\documentclass[twocolumn]{autart}    

\usepackage{graphicx}          
\usepackage{graphicx}      
\usepackage{subfigure}
\usepackage{times} 
\usepackage{amssymb}  
\usepackage{amsmath} 
\usepackage[latin1]{inputenc}
\usepackage{balance}
\usepackage{algorithm}
\usepackage[noend]{algpseudocode}
\usepackage{color}
\usepackage{natbib}

\makeatletter
\def\BState{\State\hskip-\ALG@thistlm}
\makeatother

\newcommand{\Eset}{E}
\newcommand{\Eo}{E_o}
\newcommand{\Euo}{E_{uo}}
\newcommand{\EuoM}{E_{M,uo}}
\newcommand{\Eoa}{E_{a,o}}
\newcommand{\EoM}{E_{M,o}}
\newcommand{\Euoa}{E_{a,uo}}
\newcommand{\Ec}{E_{c}}
\newcommand{\Euc}{E_{uc}}
\newcommand{\Euca}{E_{a,uc}}
\newcommand{\EucM}{E_{M,uc}}
\newcommand{\Eca}{E_{a,c}}
\newcommand{\Ecv}{E_{c,v}}
\newcommand{\Ea}{E_{a}}
\newcommand{\Ef}{E_f}
\newcommand{\Er}{E_R}

\newcommand{\Po}{P_{o}}

\newcommand{\ev}{\sigma}
\newcommand{\Xf}{X_f}
\newcommand{\Xfm}{X_f^M}
\newcommand{\Ecva}{E_{c,v}^a}
\renewcommand{\epsilon}{\varepsilon}
\newcommand{\dotcup}{\dot{\cup}}
\newcommand{\Kbar}{\overline{K}}

\newcommand{\Eov}{E_{o,v}}
\newcommand{\Eovi}{E_{o,v}^i}
\newcommand{\EM}{E_M}
\newcommand{\Eova}{E_{o,v}^a}
\newcommand{\cal}{\mathcal}

\newtheorem{definition}{Definition}
\newtheorem{remark}{Remark}
\newtheorem{example}{Example}
\newtheorem{theorem}{Theorem}

\newtheorem{proposition}{Proposition}

\def\proof{\textit{Proof: }}
\def\QED{\mbox{\rule[0pt]{1.2ex}{1.2ex}}}
\def\endproof{\hspace*{\fill}~\QED\par\endtrivlist\unskip}

\begin{document}
	
	\begin{frontmatter}


		\title{Detection and Mitigation of Classes of Attacks in Supervisory Control Systems\thanksref{footnoteinfo}} 
		
		\thanks[footnoteinfo]{This work was partially supported by the U.S.\ National Science Foundation (grant CNS-1421122) and by Brazil's CNPq (National Council of Technological and Scientific Development).}	
				
		\author[Lilian]{Lilian Kawakami Carvalho}\ead{lilian@dee.ufrj.br},    
		\author[Yi-Chin]{Yi-Chin Wu}\ead{yichin.wu@berkeley.edu},               
		\author[Raymond]{Raymond Kwong}\ead{kwong@control.utoronto.ca},  
		\author[Stephane]{St{\'e}phane Lafortune}\ead{stephane@umich.edu}  
		
		\address[Lilian]{Department of Electrical Engineering, Universidade Federal do Rio de Janeiro, Brasil}  
		\address[Yi-Chin]{Department of EECS, University of Michigan and Department of EECS, University of California at Berkeley, USA}             
		\address[Raymond]{Department of ECE, University of Toronto, Canada}        
		\address[Stephane]{Department of EECS, University of Michigan, USA}        


		\begin{abstract} 
			The deployment of control systems with network-connected components has made feedback control systems vulnerable to attacks over the network.
			This paper considers the problem of intrusion detection and mitigation in supervisory control systems, where the attacker has the ability to enable or disable vulnerable actuator commands and erase or insert vulnerable sensor readings.
			We present a mathematical model for the system under certain classes of actuator enablement attacks, sensor erasure attacks, or sensor insertion attacks. We then propose a defense strategy that aims to detect such attacks online and disables all controllable events after an attack is detected.
			We develop an algorithmic procedure for verifying whether the system can prevent damage from the attacks considered with the proposed defense strategy, where damage is modeled as the reachability of a pre-defined set of unsafe system states.
			The technical condition of interest that is necessary and sufficient in this context, termed ``GF-safe controllability'', is characterized. We show that the verification of GF-safe controllability can be performed using diagnoser or verifier automata.
			Finally, we illustrate the methodology with a traffic control system example.
		\end{abstract}
		
	\end{frontmatter}

\section{Introduction} 


The increasing amount of networked components in feedback control systems has made these systems vulnerable to cyber threats.
Since control systems are often safety critical (e.g., avionics, power grid), it is imperative to embed defense mechanisms into them \citep{Cardenas:2008,Banerjee:2012}.

\begin{figure}[tb]
	\centering
	\includegraphics[width=0.35\textwidth]{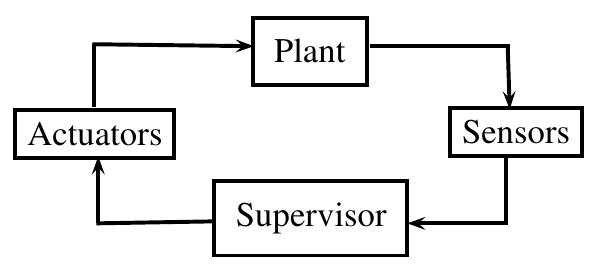}
	\caption{\label{fig:control} The closed-loop control system architecture}
\end{figure}

In this paper, we consider the closed-loop control system architecture of Figure \ref{fig:control}, where the plant is controlled by the supervisor through sensors and actuators in the traditional feedback loop.
The communication channels for the sensor and actuator signals are often unprotected, allowing attackers to potentially inject false sensor or actuator signals.

We consider event-driven supervisory control systems where the plant is abstracted as a discrete event system.
The supervisor monitors the plant behavior through the events generated by the sensors and it dynamically issues enable/disable actuator commands in order to enforce a given specification.
We study the problem of intrusion detection and mitigation for control systems under four classes of attacks: 
\emph{Actuator Enablement attacks} (AE-attacks), \emph{Actuator Disablement attacks} (AD-attacks), \emph{Sensor Erasure attacks} (SE-attacks) and \emph{Sensor Insertion attacks} (SI-attacks).
Specifically, in an attack scenario, some actuators or sensors are deemed vulnerable and the attacker can change the actuator commands (from disable to enable or vice-versa) or change the sensor readings (by erasing a genuine sensor event or inserting a fictitious one).
We address the problem of protecting the system from reaching a pre-defined set of unsafe states under each of the above attack scenarios.
Note that in general actuator attacks or sensor erasure attacks are not directly observable, while inserted fictitious sensor events are assumed to be indistinguishable from genuine ones for the supervisor.
We leverage results from supervisory control and fault diagnosis of discrete event systems and propose a defense strategy that detects attacks online and disables all controllable actuator events after detecting an attack with certainty.
This defense strategy may not be sufficient in general to prevent damage.
Hence, we characterize a property termed \emph{General Form of safe controllability} (GF-safe controllability for short) that precisely captures the capability of preventing the system from reaching an unsafe state after an attack, using the proposed defense strategy.
Here, GF stands for AE, SE, or SI.
An algorithmic procedure is developed to verify whether the system is GF-safe controllable.
For this purpose, diagnoser or verifier automata can be employed.
%

The key feature distinguishing this work from the large amount of work in cybersecurity is our focus on closed-loop control systems. 
We adopt a model-based approach to precisely capture the vulnerabilities and the effects of an attack on the control system.
The model-based approach enables a formal characterization of the unsafe behavior that an attacker tries to induce and the resiliency that the system defender wants to achieve.
The model-based approach also allows for monitoring deviations from the normal system behavior.
Our work is complementary to the works on anomaly/intrusion detection in cyber systems (e.g., \cite{Lazarevic:2005,Hoffman:2009,Zhou:2010,Modi:2013}) where detection is based on statistical analysis of network packets, for instance.
We do not focus on how attackers infiltrate vulnerable actuators or sensors, but rather on the detection of attacks and on the modeling of their effects on the control system.
Under each of the four types of attacks considered, we adopt a fairly simple attack model which can be paraphrased as ``attack whenever possible''.
However, our methodology is general and more sophisticated attack models could be embedded in it.
Similarly, our defense strategy upon detection of attacks is based on ``safety first'', by switching to a ``safe mode'' of operation, but more refined defense mechanisms could be embedded in our modeling methodology, if so desired.

Intrusion detection and prevention in the setting of supervisory control of discrete event systems have been previously studied in \cite{Thorsley:2006}, where the authors consider the design of a supervisor that achieves the specification both in normal operation and after an attack.
The focus in \cite{Thorsley:2006} 
is on finding language conditions under which the supervisor can prevent unsafe behavior in the presence of attacks while achieving a given specification, using a notion called \emph{disable language}, which shares several similarities with the safe controllability condition used in this paper.
Our focus is more explicit than \cite{Thorsley:2006} in terms of modeling several classes of attacks, detecting them algorithmically using diagnoser automata, and switching to safe mode upon detection.
The problem of intrusion detection and prevention is related to fault tolerant supervisory control problems, a well-studied problem in the literature (see, e.g., \cite{Rohloff2005,Nke2011,Paoli:2011,Sulek2014,Wen2014,Moor2015}), where a robust supervisor is designed to maintain the specification even when the system becomes faulty.
Our approach is closest to the work in \cite{Paoli:2011}, where the authors consider a strategy that detects faults online and reconfigures the control law when a fault is detected.
Our notion of GF-safe controllability is a GF-attack variant of the \emph{safe controllability} property introduced  in \cite{Paoli:2011}.

The main contributions of this paper are as follows.
First, we present a mathematical model for supervisory control systems under AE-attacks and propose a defense strategy that detects attacks online and, upon detection with certainty, disables all controllable events in order to prevent attack damage.
We define the property of \emph{AE-safe controllability} that characterizes the system's capability to prevent damage under AE-attacks and develop algorithmic procedures for verifying AE-safe controllability using diagnoser and verifier automata. 
Next, we consider other types of attacks.
We only briefly discuss AD-attacks and focus instead on SE- and SI-attacks. 
Paralleling the case of AE-attacks, we model the effect of SE- and SI-attacks on the control system.
{For AE- and SE- attacks, we consider a worst-case scenario where the attacker may attack at every opportunity.
For SI-attacks, we consider an attack strategy where the attacker never inserts a sensor reading that is not defined in the current state of the nominal supervisor.}
We then generalize AE-safe controllability to GF-safe controllability, the property that the system should satisfy in order to successfully prevent damage from either AE-, SE- or SI-attacks, and finally we develop a test to verify GF-safe controllability. 
In the case of SE- and SI-attacks, in addition to testing the corresponding version of GF-safe-controllability, it is also necessary to test if the control system under attack has a deadlock. 

%

The remainder of this paper is organized as follows. 
We define the types of attacks we deal with in Section~\ref{sec:types_attacks}.
Section~\ref{sec:back} introduces our mathematical framework.
Section~\ref{sec:aea} studies the effect of actuator enablement attacks.
Then, in Section~\ref{sec:det_mit_attacks}, we define the property of AE-safe controllability and discuss its verification.
We present the model of the system under sensor erasure and insertion attacks in Sections~\ref{sec:se} and \ref{sec:si}, respectively. 
In Section~\ref{sec:GF}, we define the property of GF-safe controllability and present an algorithm for its verification.
Finally, in Section \ref{sec:exemple}, we illustrate our methodology with a traffic control system example and in Section \ref{sec:conclusion}, we conclude the paper. 

A preliminary and partial version of the results in Sections 4 and 5 was presented in \cite{Carvalho:2016}. 
The results in Sections 5.4, 6, 7, and 8 are new.

\section{Types of attacks} 
\label{sec:types_attacks}
\begin{figure}[thpb]
	\centering
	\includegraphics[width=0.35\textwidth]{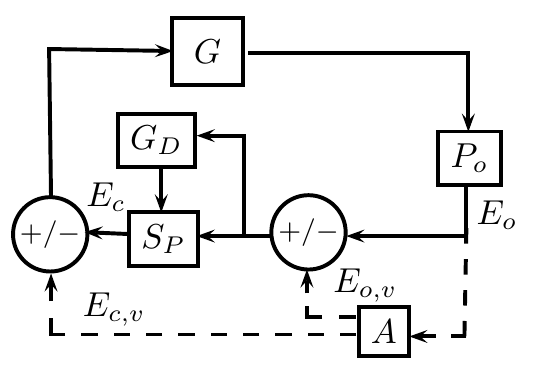}
	\caption{\label{fig:attacks} The control system architecture under attack}
\end{figure}

We depict in Figure \ref{fig:attacks} the attack model under consideration. 
The control system architecture under attack has a plant $G$ equipped with a set of
potentially vulnerable sensors and actuators, and $G$ is controlled by a partial-observation supervisor (or P-supervisor) $S_P$.
Let $E$ be the event set of $G$.
The actuators are modeled by the set of \emph{controllable} events $E_c$, with $E_c \subseteq E$, while the sensors are modeled by the set of \emph{observable} events $E_o$, with $E_o \subseteq E$.
The supervisor observes the occurrences of the plant's observable events through projection $P_o$ from set $E$ to set $E_o$. 
The attacker, represented by block $A$, has access to subsets of $E_c$ and $E_o$, representing \emph{vulnerable} actuators and sensors and denoted by $E_{c,v} \subseteq E_c$ and $E_{o,v} \subseteq E_o$, respectively.
The sets $E_{c,v}$ and $E_{o,v}$ are predefined based on system knowledge and are application dependent. They can, for example, reflect the capability of the attacker to exploit vulnerabilities of the system.
Finally, block $G_D$ is the module that detects attacks, which we call the \emph{intrusion detection module}.

The fact that the attacker can compromise either sensors or actuators is captured by the two outputs of $A$ that affect:
 (i) the actual observations of $S_P$ and $G_D$, which consist of the genuine sensor readings affected by the attacks on them; 
 and (ii) the actual control actions that are applied to $G$, which consist of the combination of the genuine control actions of $S_P$ with those of $A$. 
The combination of the attacks of $A$ with genuine sensor readings and genuine control actions are denoted by the two $+/-$ blocks in Figure \ref{fig:attacks}. This $+/-$ is a conceptual operation and represents the following four types of attack of $A$ considered herein:
\begin{description}
	\item[AE] for Actuator Enablement: $A$ ``overrides" a control action of $S_P$ on a particular controllable event in $E_{c,v}$, by enabling an event that is currently disabled by $S_P$;
	\item[AD] for Actuator Disablement: $A$ ``overrides" a control action of $S_P$ on a particular controllable event in $E_{c,v}$, by disabling an event that is currently enabled by $S_P$;
	\item[SE] for Sensor Erasure: $A$ ``erases" an occurrence of an observable event in $E_{o,v}$, thereby making that occurrence unobservable to $S_P$ and $G_D$;
	\item[SI] for Sensor Insertion: $A$ ``inserts" a fictitious occurrence of an observable event  in $E_{o,v}$ to the observation stream of $S_P$ and $G_D$.
\end{description}
{Hereafter, we assume that $S_P$ has already been designed and is fixed.
The goal is to design $G_D$ to detect and mitigate attacks by $A$ in each of the four above cases.
The attack model of $A$ that is considered by $G_D$ will be discussed in each case.}

\section{Mathematical framework}\label{sec:back} 

We consider plants modeled as deterministic finite-state automata.
An automaton is denoted by $G=(X,E,f,x_0,X_m)$, where $X$ is the finite set of states, $\Eset$ is the finite set of events, $f: X \times E^* \rightarrow X$ is the (potentially partial) transition function, $x_0$ is the initial state, and $X_m$ is the set of marked states. 
For the sake of simplicity, the set of marked states will be omitted unless blocking is considered.
The language generated by $G$ is the set of strings defined by ${\mathcal L}(G):= \{s \in E^*: f(x_0, s) \mbox{~is defined}\}$
and
the language marked by $G$ is  ${\mathcal L}_m(G):= \{s \in E^*: f(x_0, s) \in X_m\}$.
Consider event set $E' \subseteq \Eset$ and state $x \in X$;
the set of reachable states with respect to $E'$ and $x$ is defined as
$Reach(G,x,E')=\{x' \in X: (\exists s \in E'^*)[f(x,s) = x']\}$.
The active event set of $G$ at state $x$ is denoted by $\Gamma_G(x)$.

As was mentioned above, $\Eset$ is partitioned as $\Eset = \Eo \dotcup \Euo$, where $\Eo$ and $\Euo$ denote, respectively, the sets of observable and unobservable events;
similarly, $\Eset = \Ec \dotcup \Euc$, where $\Ec$ and $\Euc$ denote, respectively, the sets of controllable and uncontrollable events.
When an event $\ev$ appears in string $s$, we write $\ev \in s$.
Similarly, we write $\Ec \in s$ whenever $s$ has an event in $\Ec$.
The natural projection $P_o : \Eset^* \rightarrow  \Eo^*$ is defined such that {\em (i)\/} $P_o(\epsilon) =  \epsilon$; {\em (ii)\/} $P_o(\ev) = \ev$ if $\ev \in \Eo$; {\em (iii)\/} $P_o(\ev) = \epsilon$ if $\ev \in \Euo$; and {\em (iv)\/} $P_o(s \ev) = P_o(s) P_o(\ev)$ for $s \in \Eset^*$ and $\ev \in \Eset$, where $\epsilon$ denotes the empty string. 
Given $s \in E_o^*$, the inverse projection of $t$ is $\Po^{-1}(t)= \{ s \in E^* : P_o(s) = t \}$. 
Both the projection and the inverse projection operations are extended to languages by applying $P_o(s)$ and $\Po^{-1}(s)$ to all strings in the language.
We write $s'<  s$ when $s'$ is a strict prefix of $s$.
Given $L \subseteq E^*$, we define $L/s:= \{t : st \in L \}$, which is the set of all suffix strings in $L$ after $s$.

When it is necessary to restrict the behavior of $G$ in order to satisfy some performance specification $K \subseteq {\mathcal L}(G)$, we introduce a feedback control loop together with a \emph{supervisor}.
We consider specifications defined in terms of admissible sublanguages of ${\cal L}(G)$.
The supervisor dynamically enables or disables events of the plant \citep{Ramadge:1989}, restricting the closed-loop behavior within the admissible language.
In general, the plant is partially observable and thus the supervisor decides which events to be disabled based on the projections of strings generated by $G$. 
More specifically, a supervisor under partial observation is a mapping  $S_P : \Po[{\cal L}(G)]  \rightarrow  2^\Eset$; for every string $s$ generated by $G$, the supervisor makes its decision based on $P_o(s)$. 
As a consequence, two different strings $s_1$ and $s_2$ with the same projection lead to the same control action. 
Such a supervisor is referred to as a P-supervisor, and the resulting controlled system is denoted by $S_P/G$. 

We say that sublanguage $K$ of ${\cal L}(G)$ is \emph{controllable} with respect to ${\cal L}(G)$ and $\Euc$ if $\Kbar \Euc \cap {\cal L}(G) \subseteq \Kbar$.
Also, $K$ is \emph{observable} with respect to ${\cal L}(G)$, $\Po$ and $\Ec$ if for all  $s \in \Kbar$ and $\ev \in \Ec$, $s \ev \notin \Kbar$ and $s \ev \in {\cal L}(G)$ implies that $\Po^{-1}[\Po(s)]\ev \cap \Kbar= \emptyset$.
It is well-known that controllability and observability are necessary and sufficient for the existence of a supervisor that enforces $\overline{K}$ \citep{Wonham:2013}.

\section{Actuator enablement attacks}\label{sec:aea} 
This section and the next one consider a supervisory control system with vulnerable actuators.
Specifically, we consider an attack scenario where the attacker has infiltrated a set of vulnerable actuators and overrides ``disable'' control actions from the supervisor by ``enable'' actions for the compromised actuators. 
The goal of the attacker is to use these ``enable'' control actions to potentially drive the system to an unsafe state. 
We call such attacks \emph{Actuator Enablement attacks}, or \emph{AE-attacks} for short.

To represent the AE-attack model in Figure \ref{fig:attacks}, the combination of the control actions of the supervisor $S_P$ and the attacker $A$ ($+/-$ block) is to be interpreted as the OR operation on the control actions (i.e., enabled events) of $S_P$ and those of the attacker $A$. 
{Recall that the set of vulnerable actuator events is denoted by $\Ecv$, which is a subset of $\Ec$.
The vulnerable actuator events in $\Ecv$ can be either observable or unobservable.
Our methodology  accounts for both cases.}

The attacker potentially observes the same set of observable events through $P_o$ as the system does (this is left unspecified), and it
can override the supervisor's control actions on vulnerable events. 
Ignoring attacks, the closed-loop behavior is ${\cal L}(S_P/G)=\overline{K}$, where $K$ is a controllable and observable sublanguage of ${\cal L}(G)$.
That is, $S_P$ is the ``nominal'' supervisor that was designed to enforce $K$.
It may or may not be resilient to attacks; this is what we wish to determine.

{Module $G_D$ receives the occurrences of observable events through projection $P_o$ and its goal is to infer the presence of AE-attacks.
When such a detection occurs with certainty, we adopt the simple defense model that $G_D$ forces $S_P$ to switch from enforcing $K$ to a \emph{safe mode}, where all controllable events are permanently disabled.
In the development that follows, we assume that $G_D$ has no prior knowledge of the attack model of $A$, so $G_D$ will consider that $A$ can potentially override every disable command to a vulnerable actuator; in other words, $G_D$ assumes a worst-case attack scenario.
But other attack scenarios could be considered by suitably altering the modeling methodology presented next.

The simple defense strategy of disabling all controllable events corresponds to ``expect the worst and put safety first''. Our primary focus in this paper is to develop a precise model for various types of attacks in supervisory control systems and to understand the effects of such attacks. This problem does not appear to have been studied in this formal manner in the literature. Since this is the objective of this work, we have adopted the  simple and conservative ``safety first'' approach to defend attacks, and have left the refinement of our methodology to account for more sophisticated defense mechanisms, as well as other issues such as blocking, for future work.



%
%



We now describe how to model the closed-loop system under the above scenario of an AE-attack;
then we will show how to design the intrusion detection module $G_D$ in Section \ref{sec:det_mit_attacks}.
We employ two operations in our modeling methodology: \emph{dilation} and \emph{compression} \citep{Carvalho:2012,Alves:2014}. 
These operations are useful for modeling the  attacker's actions. In order to do so, let $\Ecva=\{\ev^a : \ev \in \Ecv\}$ denote the set of attacker's events on vulnerable actuators, which we will refer to as \emph{attacked actuator events} and define $\Ea = \Eset \cup \Ecva$.
The dilation operation is a mapping $ D :  \Eset^*  \rightarrow 2^{\Ea^*}$ with the following properties: {\em (i)\/} $D(\epsilon)  =  \{ \epsilon \}$; {\em (ii)\/} $D(\ev)  = \{\ev\}$ if $ \ev \in \Eset\backslash \Ecv$; {\em (iii)\/} $D(\ev)  = \{\ev, \ev^a\}$ if  $\ev \in \Ecv$; and {\em (iv)\/} $D(s \ev) =  D(s) D(\ev)$ where $s\in \Eset^*$ and $\ev \in \Eset$.
The compression operation recovers a string $s$ from a dilation string in $\Ea^*$. 
It is a mapping $C:  \Ea^\ast  \rightarrow  \Eset^\ast$ such that {\em (i)\/} $C(\epsilon) = \epsilon$; {\em (ii)\/} $C(\ev) = \ev$, if $\ev \in \Eset$; {\em (iii)\/}  $C(\ev^a) = \ev$, if $\ev^a \in \Ecva$; and {\em (iv)\/} $C(s_a\ev) = C(s_a)C(\ev)$ where $s_a \in \Ea^*$ and $\ev \in \Ea$.
Both the dilation and the compression operations can be extended to languages by applying them to all strings in the language.
That is, $D(L)=\bigcup_{s\in L}D(s)$ and $C(L_{a}) = \bigcup_{s_a \in L_{a}}C(s_a)$.

We present in Algorithm \ref{alg:ae} the construction of the closed-loop system under AE-attacks. 
Consider the plant $G$ and let $H$ be the finite-state automaton realization of supervisor $S_P$.
Recall that the realization of a partial-observation supervisor captures in its active event set the current set of enabled events;
in particular, enabled unobservable events are captured by self-loops at the current state of $H$.

First, we construct $G_a$ by adding to $G$ all possible attacker actions using the dilation operator $D$ on ${\mathcal L}(G)$. 
For a transition labeled by $\ev \in \Ecv$ on $G$, we add in parallel a transition labeled by $\ev^a$ to represent an AE-attack.
This captures an attack by $A$ on each transition representing a vulnerable actuator event.

Next, we build $H_a$, the overall supervisor under the effect of AE-attacks.
Specifically, we take the supervisor realization $H$ and add self-loops to all of its states with events in $\Ecva$, \emph{when the compression of the candidate event is not in the active event set of the state}.
These self-loops for attack events model the attacker's ability to enable attacked actuator events, when those events are disabled by $S_P$. 
{In addition, to capture the fact that a supervisor should never disable an uncontrollable system event, we also add self-loops for every uncontrollable event, when these events are not in the active event set of the state. 
Indeed, after an AE-attack, new occurrences of uncontrollable events could occur that are not defined at the current supervisor state (since the plant may have changed state {unknown to the supervisor} due to an AE-attack).}

Finally, we find the closed-loop system under AE-attacks, $G_{M}$, by parallel composing $H_a$ and $G_a$. 
Automaton $G_{M}$ models the behavior of the system in the presence of AE-attacks on \emph{all} vulnerable actuators at \emph{all} times, which corresponds to the worst-case scenario under consideration.
For simplicity, in the remainder of this paper, we will write $L_M$ for ${\cal L}(G_M)$.
Clearly, by construction of $G_M$, $L_M$ will be a controllable and observable sublanguage of ${\mathcal L}(G_a)$.



\begin{algorithm}[htb]\caption{Algorithm for AE-attack model}\label{alg:ae}
	\begin{algorithmic}[1]
		\Statex \textbf{Inputs:}
		\begin{itemize} \item $G=(X,E,f,x_0)$ and $H=(X_{H},\Eset,f_{H},x_{0,{H}})$ : plant and supervisor realizations, respectively \item   $\Eo$, $\Ec$ and $\Ecv$ : sets of observable, controllable and vulnerable actuator events \end{itemize} 					
		\Statex \textbf{Output:} 
		Closed-loop system under AE-attacks $G_{M}=(X_M,\Ea,f_M,x_{0,M})$
		\State Build $G_a=(X,E_{a},f_{a},x_{0})$, where
$f_a(x,\ev^a):=f(x,C(\ev^a))$ if $f(x,C(\ev^a))$ is defined, $\forall \ev^a \in \Ea $, $\forall x \in X$
		\State Build $H_a=(X_{H},E_a,f_{H_a},x_{0,{H}})$, where
		\begin{align*}
		&f_{H_a}(x_H,\ev) =\\&
		\begin{cases}
			f_H(x_H,\ev), &\mbox{if $f_H(x_H,\ev)$ is defined}\\
			x_H, &\mbox{if ($\ev \in \Ecva\wedge f_H(x_H,C(\ev))$}\\
				 &\mbox{is undefined) $\vee$ ($\ev \in {\Euc} \wedge$} \\
			     &\mbox{$f_H(x_H,\ev)$ is undefined)}
		\end{cases}
		\end{align*}	
		\State Compute $G_{M}=H_a\|G_a$
	\end{algorithmic}
\end{algorithm}

In $G_M$, the only controllable events are those in $E_c$, since the events in  $E_{c,v}^a$ are actions of the attacker and thus uncontrollable.
Note that the events in $E_{c,v}$ are indeed controllable, but of course they can be overridden by the corresponding events in $E_{c,v}^a$.
Also, the observability properties of the events in $E_{c,v}^a$ are inherited from the corresponding ones in $E_c$.

\begin{example}
	\label{ex:ex1}
	We consider the plant $G$ in Figure \ref{fig:ex1_G} with $\Ec=\Ecv=\{b\}$. 
	State $4$ is the unsafe state of the plant and it is identifed with a square.
	The supervisor that controls $G$ is realized as automaton $H$ in Figure \ref{fig:ex1_H}.
	Notice that the supervisor disables event $b$ at state $2$, thereby preventing the plant from reaching unsafe state $4$.
	\begin{figure}[h]
		\center
		\subfigure[\label{fig:ex1_G}$G$: plant model]{\includegraphics[width=0.25\textwidth]{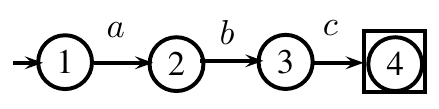}}
		\qquad
		\subfigure[\label{fig:ex1_H} $H$:  supervisor realization]{\includegraphics{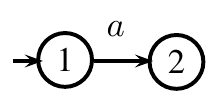}}
		
		\subfigure[\label{fig:ex1_Ga}  $G_a$:  plant subject to attack]{\includegraphics{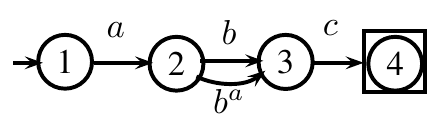}}\\
		
		\subfigure[\label{fig:ex1_HA}  $H_a$:  supervisor realization including the effects of the attack]{\includegraphics[width=0.25\textwidth]{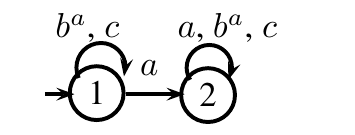}}\\
		
		\subfigure[\label{fig:ex1_Gm}  $G_M$: closed-loop system under attack]{\includegraphics[width=0.25\textwidth]{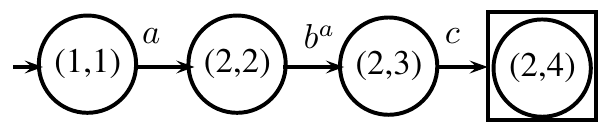}}\\

		\caption{Figures of Example \ref{ex:ex1}}
	\end{figure}
Following Algorithm \ref{alg:ae}, we build $G_a$ in Figure \ref{fig:ex1_Ga} by adding a transition labeled by $b^a$ in parallel with the transition labeled by $b$. 
We then build  in Figure \ref{fig:ex1_HA} the realization of the supervisor under AE-attacks by adding a self-loop for attacked actuator event $b^a$ at every state;
we also add self-loops for uncontrollable events $a$ and $c$ when they are not in the active event set of the state. 
Finally, we build in Figure \ref{fig:ex1_Gm} the closed-loop system under AE-attacks by $G_{M}=H_a||G_a$.
Each state in $G_{M}$ is a pair where the first state is the supervisor state and the second state is the plant state.  
We can see that, with the attacker enablement of vulnerable actuator event $b$, the plant can transition from state $2$ to state $3$ and then reach unsafe state $4$ through uncontrollable event $c$. 
%
\end{example}

\section{Detection and mitigation of actuator enablement attacks} \label{sec:det_mit_attacks} 

\subsection{Detection and mitigation strategy} 

As we can see in Example \ref{ex:ex1}, under AE-attacks, the plant can deviate from the specification enforced by the supervisor and reach an unsafe state. 
To mitigate the effects of such attacks, our strategy is to design an attack detection module and then switch to ``safe mode'' of operation when an {attack} has been detected.
This defense strategy may or may not be sufficient to prevent the system from reaching a set of states deemed \emph{unsafe}.
Our goal is to identify a condition under which this defense strategy does work.
We model the set of unsafe states distinctly from the original specification $K$ achieved by $S_P$.
That is, while all states reached by $S_P/G$ are assumed to be safe, not all states outside of those reached by $S_P/G$ may be unsafe.
We denote the set of unsafe states by $X_f$.
$X_f$ is a subset of $X$ that captures physical states where damage to the plant would occur, for instance. Such states can be determined from properties of the physical system when the automaton model is developed.

Our techniques are adapted from techniques developed in \cite{Paoli:2011} for ``safe controllability'' and in \cite{Thorsley:2006} for ``disable languages''. 
Specifically, with the model developed in the previous section, we formulate the problem of attack detection as a fault diagnosis problem, where the fault events are the attacker's actions on vulnerable actuator events.
We design an intrusion detection module that monitors the output from the plant and notifies the supervisor when an attack has been detected (with certainty).
The supervisor, upon receiving an attack report from the intrusion detection module, switches to its {safe mode} of operation where it disables \emph{all} controllable events. 
We remark that the attack detection together with the safe controllability strategy derived here are also suitable for \emph{on-the-fly}  implementations, since they rely solely on diagnosers, which can be constructed on-the-fly (as opposed to synthesized off-line).

\subsection{AE-safe controllability} 
\label{sec:AE-SC}
{We define a variant of safe controllability from \cite{Paoli:2011} in the context of AE-attacks and call it \emph{AE-Safe Controllability}; it is formally defined in Definition \ref{def:safcon} below.
	Specifically, consider the set of unsafe states $X_f \subset X$. 
	{Let $\Psi(\Ecva) = \{ t \in {\cal L}(G) : t  = t'\ev, ~t' \in E_a^*, \ev \in \Ecva \}$ denote the set of strings for which the last event is an attacked actuator event.}
%
%
	%
	Consider $G_{M}$ built in Algorithm \ref{alg:ae} that models the closed-loop system subject to AE-attacks and let $X_f^M=\{(x_H,x)\in X_M:  x\in X_f\}$ be the set of unsafe states in $G_{M}$.
	In words, AE-safe controllability holds if we can detect any attack occurrence and then disable a controllable event before the plant reaches an unsafe state.
	For the purpose of the definition that follows, we define the following projection: 
	$\Po^a: E_a^* \rightarrow(\Eo\cup D(\Ecv\cap\Eo))^*$.
	
	
{	\begin{definition}[AE-Safe Controllability]\label{def:safcon}
		Consider $G_{M}=(X_M,\Ea,f_M,x_{0,M})$ from Algorithm~\ref{alg:ae}. 
		Language $L_M= {\cal L}(G_{M})$  is AE-safe controllable with respect to projection $\Po^a$, attacked actuator events $\Ecva$, and unsafe states $X_f^M$ if	
		$(\forall s \in \Psi(\Ecva))(\forall t \in L_M/s)$ $[(f_M(x_o,st) \cap X_f^M \neq \emptyset) \wedge (\forall s'<  st, f_M(x_o,s') \cap X_f^M = \emptyset)] \Rightarrow$
		$(\exists t_1,t_2 \in E_a^*) [(t = t_1 t_2) \wedge \left((\nexists \omega \in L_M)[\Po^a(st_1)=\Po^a(\omega) \wedge \Ecva \notin \omega]\right)\wedge (\Ec \in t_2)]$.
	\end{definition}
}
	\begin{figure}[htb]
		\centering
		\includegraphics[width=0.35\textwidth]{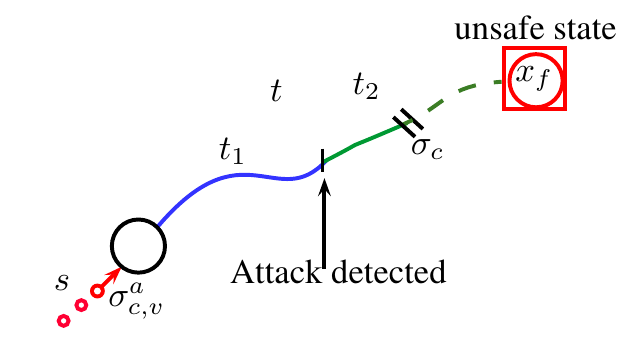}
		\caption{\label{fig:SF} Illustration of AE-safe controllability where $\ev_{c,v}^a \in \Ecva$, $t=t_1t_2$, $\ev_c \in \Ec$, and $x_f \in X_f^M$}
	\end{figure}
	We will sometimes slightly abuse terminology and say that ``system $G$ is AE-safe controllable'' if the corresponding
	$L_M$, $P_o^a$, $\Ecva$ and $X_f^M$ are understood and if Definition~\ref{def:safcon} holds.
	Figure \ref{fig:SF} illustrates the definition of AE-safe controllability. 
	The first state is reached through a string $s$ whose last event $\ev_{c,v}^a$ is an attacked actuator event. 
	String $t$ is the continuation of $s$ that reaches an unsafe state for the first time. 
AE-safe controllable holds if for every such $s$ and $t$, $t$ can be written as $t=t_1t_2$ where (1) the attacked actuator event can be detected after $st_1$ and (2) $t_2$ contains a controllable event in $E_c$.
	Recall that all events in $E_c$ are controllable and that events in $\Ecva$ are uncontrollable in $G_M$.
	That is, AE-safe controllability holds if we can detect an attack occurrence and then disable a controllable event before the plant reaches an unsafe state; and this property has to hold for every attack occurrence.
	It should be noted that the detection condition after string $st_1$ is that an attack has been detected on any of the vulnerable actuators (cf.\ $\Ecva \notin \omega$ in detection clause in definition), not necessarily for the same event at the end of string $s$; as long as module $G_D$ knows for sure that one vulnerable actuator was indeed attacked, then it forces $S_P$ to switch to safe mode.
	
	
	The construction procedure of $G_M$  and the conditions in the definition of AE-safe controllability lead directly to the following result, whose proof is omitted.
	\begin{theorem}\label{thm:one}
		Under the attack and defense model considered in this paper, system $G$ will not reach an unsafe state if and only if it is AE-safe controllable w.r.t.\ projection $\Po^a$, vulnerable actuator events $\Ecv$, and set of unsafe states $X_f^M$.
	\end{theorem}
}
\subsection{Test of AE-safe controllability using diagnoser}\label{sec:diagnoser} 


To test if a system is AE-safe controllable, we develop an algorithmic procedure that relies on diagnoser automata (or simply, diagnosers).
The diagnoser, as developed in \cite{Sampath:1995}, relies on the computation of the observer of the automaton obtained by performing a parallel composition between the plant automaton and the so-called label automaton that captures occurrences of faults, as described in \cite{cassandras2008}.    
Our algorithm verifies if the diagnoser-based intrusion detection module can detect any attack before the plant reaches an unsafe state and if the supervisor can disable events to prevent the plant from reaching $X_f$. 
Before we formally present the algorithm, we first review the definition of the set of first-entered certain states in a diagnoser from \cite{Paoli:2011}; the reader is referred to \cite{cassandras2008} for the definition of diagnoser and any undefined terminology.

\begin{definition}[Set of first-entered certain states] \label{def:FCS}	
Let $G_d = (Q_d, E_o, f_d, q_{0,d})$ be the diagnoser constructed from a given plant and the appropriate label automaton. 
Define $Q_{YN}=\{q \in Q_d:q \text{ is uncertain}\}$, $Q_{N}=\{q \in Q_d: q \text{ is normal}\}$, and $Q_{Y}=\{q \in Q_d: q \text{ is certain}\}$.
The set of first-entered certain sates is  ${\cal FC} = \{q \in Q_Y: (\exists q'\in Q_{YN}\cup Q_{N}, \exists \ev \in \Eo)[f_d(q',\ev)=q]\}$.
\end{definition}

{We can now present Algorithm~\ref{alg:sc}, the diagnoser-based algorithm for testing AE-safe controllability.
By construction of $G_M$, we can see that our goal is to detect occurrences of events in $\Ecva$ in $L_M$, based on observable event set $\Eoa$;
specifically, the attacked actuator events in $\Ecva$ are the ``fault'' events to be diagnosed, and they are assumed to {be all} of the same fault type.}
Hence, we wish to build the diagnoser of $G_M$.
In step 1, we consider the label automaton $A_\ell$ in Figure \ref{fig:label-automaton} and label the attacked actuator events $\Ecva$ by building $G_\ell= G_{M} || A_\ell$.
\begin{figure}[htb]
	\centering
	\includegraphics[width=0.25\textwidth]{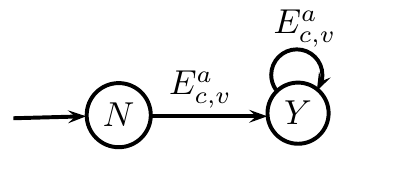}
	\caption{\label{fig:label-automaton} Label automaton $A_\ell$}
\end{figure}
We then compute in step 2 the diagnoser automaton $G_d=Obs(G_\ell,\Euoa )$, where  $Obs(G_\ell,\Euoa )$ denotes the observer of $G_\ell$ with respect to unobservable event set $\Euoa$, where $\Euoa=\Euo\cup D(\Ecv\cap\Euo)$.
In step 3, we test if any uncertain state contains an unsafe state. 
If this is the case, then the diagnoser cannot detect the attack before the plant reaches an unsafe state; hence, AE-safe controllability is violated. 
Next, we compute the set of first-entered certain states ${\cal FC}$ and then verify in step 6 if any state in $\mathcal{FC}$ contains an unsafe state.
If this happens, then even though the attack is detected, it already caused the plant to reach an unsafe state;
hence, the system is not  AE-safe controllable. 
Finally, we find the set of states reachable from $\mathcal{FC}$  through uncontrollable or attacked actuator events, and then test in step 10 whether this set contains any unsafe state.
If this happens, then even though the attack has been detected, the plant can still uncontrollably reach an unsafe state and is therefore not AE-safe controllable.
In the algorithm, $q_{\downarrow x} := \{x: (\exists l)[(x, l) \in q]\}$ is the projection of $q$ to the set of corresponding $G_M$ states.

\begin{algorithm}[htb]
	\caption{AE-safe controllability test using diagnoser}
	\label{alg:sc}
	\begin{algorithmic}[1]
		\Statex \textbf{Inputs:}\begin{itemize} \item $G_{M}=(X_M,\Ea,f_M,x_{0,M})$: closed-loop system subject
			to AE-attacks 
			\item $\Xf$: set of unsafe states \item $\Ecva$ : set of attacked actuator events\end{itemize}
		\Statex \textbf{Output:} AESafeControllability $\in \{true,false\}$
		\State Build $G_\ell= G_{M} || A_\ell$, where $A_\ell$ is shown in Figure \ref{fig:label-automaton}
		\State Compute diagnoser $G_d=Obs(G_\ell,\Euoa )$, where $\Euoa=\Euo\cup D(\Ecv\cap\Euo)$
		\If {there is uncertain state $q=\{(x_{i_1},\ell_{i_1}),\ldots, (x_{i_n},\ell_{i_n})\}$\Statex{$\in Q_{YN} $ in which there exists $x_{i_j} \in \Xfm$}}
			\State{AESafeControllability $=$ false}
		\Else{
			Compute $\cal{FC}$ according to Definition \ref{def:FCS}
			\If {there is $q=\{(x_{i_1},Y), \ldots, (x_{i_n},Y)\}\in \mathcal{FC} $  in which there exists $x_{i_j} \in \Xfm$} 
				\State AESafeControllability $=$ false
			\Else{
				 \State Compute \small{$$X^{uc}=\bigcup_{q \in \mathcal{FC}} \bigcup_{x_{M}\in q_{\downarrow x}} Reach(G_{M},x_{M},\Euc\cup\Ecva)$$}           
				\If {$ X^{uc} \cap \Xfm \neq \emptyset$ } 
					\State AESafeControllability $=$ false
				\Else { AESafeControllability $=$ true}
				\EndIf}
			\EndIf}
		\EndIf
	\end{algorithmic}
\end{algorithm}
\begin{proposition}\label{thm:diag}
Consider $G_{M}=(X_M,\Ea,f_M,x_{0,M})$ from Algorithm~\ref{alg:ae}. 
Automaton $G_d$ is the diagnoser built in Algorithm \ref{alg:sc}. 
	Language $L_M$ is \emph{not} AE-safe controllable with respect to $\Po^a$, $\Ecva$, and $\Xfm$ if and only if one of the following conditions holds true:
	\begin{enumerate}
		\item There exists $q_{YN}=\{(x_{i_1},\ell_{i_1}), \ldots, (x_{i_n},\ell_{i_n})\}\in Q_{YN} $ 
		such that $\exists j \in \{1,\ldots, n\}$, $x_{i_j} \in \Xfm$ and $\ell_{i_j}=Y$.
		\item There exists $q_{Y}=\{(x_{i_1},Y), \ldots, (x_{i_n},Y)\}\in \mathcal{FC} $ 
		such that $\exists j \in \{1,\ldots ,n\}$, $x_{i_j} \in \Xfm$.
		\item There exists $x_M \in X^{uc}$ such that $x_M \in \Xfm$, where $X^{uc}$ is defined in Algorithm \ref{alg:sc}.
	\end{enumerate}
\end{proposition}

\proof Given in Appendix.

{Note that the diagnoser will always immediately detect the attacks on vulnerable events in $\sigma \in \Ecv \cap \Eo$, since the corresponding event $\sigma_a$ is observable. 
However, in this case, the plant may still reach an unsafe state via uncontrollable and attacked actuator events, violating AE-safe controllability.
Hence, the conditions in Definition~\ref{def:safcon} still need to be tested, as described in Algorithm~\ref{alg:sc}.}

\begin{example}
	Returning to Example \ref{ex:ex1}, we show the closed-loop system under AE-attacks in Figure \ref{fig:ex1_Gm}. 
	We follow Algorithm \ref{alg:sc} to test whether the system is AE-safe controllable.
	In step 1, we build $G_\ell$ with respect to $\Ecva= \{b^a\}$ in Figure \ref{fig:ex1_Gl}. 
	Assuming $\Euoa=\emptyset$ for simplicity, the diagnoser is the same automaton as $G_\ell$. 
	By examining the diagnoser states in Figure \ref{fig:ex1_Gl}, we see that the attack will be detected in diagnoser state $((2,3), Y)$, before the plant reaches unsafe state $4$.
	However, with the test in step 10, we find that $X^{uc} = \{(2,3),(2,4)\}$ contains unsafe state $(2,4)$.
	That is, although the diagnoser can detect the attack before entering an unsafe state, since the supervisor cannot disable uncontrollable event $c$, the plant can still reach unsafe state $(2,4) \in \Xfm$ under attack.	
	Consequently, AE-safe {controllability} is violated.

	\begin{figure}[htb]
		\centering
		\includegraphics{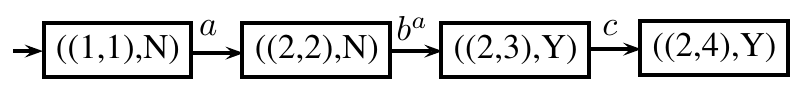}
		\caption{\label{fig:ex1_Gl} Automaton $G_\ell$}
	\end{figure}
\end{example}


{
\subsection{Test of AE-safe controllability using verifier} 

Another way to verify language diagnosability is by using verifier automata, or simply verifiers \citep{Yoo:2002a,Jiang:2001,Moreira:2011}. The main advantage of verifiers over diagnosers is that their computation requires polynomial time in the state space of the automaton, while building diagnosers will have complexity exponential in the number of states of the plant automaton in the worst case. On the other hand, unlike diagnosers, verifiers are not suitable for online diagnosis. 

Algorithm~\ref{alg:scv} tests AE-safe controllability using a verifier.
Step~1 of Algorithm~\ref{alg:scv} is the same as step 1 of Algorithm \ref{alg:sc}.
In step~2, verifier $G_V$ is built based on the methodology in \cite{Moreira:2011} (which is only briefly reviewed here).
The construction of $G_V$ starts by computing automata $G_N$ and $G_F$ that model the normal and the faulty/attacked behavior of the system, respectively. After obtaining $G_N$ (with state space denoted by $X_N$), we rename its unobservable events using the renaming function $R:\Ea\setminus\Ecva \rightarrow \Er$, where $R(\ev)=\ev$, if $\ev \in\Eoa$ and $R(\ev)=\ev_R$, if $\ev \in \Euoa\setminus\Ecva$. Thus, the unobservable events of $G_N$ and $G_F$ become ``private'' events of these automata.
In step 3, we test if any state in verifier $G_V$ is an unsafe state.
In step 5, we complete $G_V$ by adding observable events to a new state $A$. This new state marks a possible attack detection.
For state $A$, only uncontrollable events are added, since after diagnosing the attack, AE-safe controllability is violated if there exists a trace that reaches an unsafe state through unobservable events only.
In step~6, $G_T$ tracks all traces that, after the attack has been diagnosed, have only uncontrollable events in their continuations;
its state space is denoted by $X_T$.
In step 7, if $G_T$ contains an unsafe state, then the attack can steer the system to an unsafe sate before diagnosis of an attack.
%
\begin{algorithm}[htb]
	\caption{AE-safe controllability test using verifier}
	\label{alg:scv}
	\begin{algorithmic}[1]
		\Statex \textbf{Inputs:}\begin{itemize} \item $G_M =(X_M,\Ea,f_M,x_{0,M})$: closed-loop system subject to actuator enablement attacks \item $\Xf$: set of unsafe states \item $\Ecva$ : set of attacked actuator events \end{itemize}
		\Statex \textbf{Output:} \begin{itemize}  \item SafeControllability $\in \{true,false\}$\end{itemize}
		\State Build $G_\ell=G_M\| A_\ell$, where the label automaton $A_\ell$ is shown in Figure \ref{fig:label-automaton}
		\State Build verifier automaton $G_V=(X_V,\Er\cup\Ea,f_V,x_{0,V})$ assuming $\Ecva$ the set of fault events according to Algorithm 1 in \cite{Moreira:2011} 
		\If {there exists $\{(x_N,N),(x,Y)\}$ of $G_V$ such that $x \in \Xfm$} SafeControllability = false \Else{
			\State Build $G_V^{cd}=(X_V^{cd},\Er\cup\Ea,f^{cd}_{V},x_{0,{V}})$, where
			\begin{itemize}
				\item $X_V^{cd} = X_V \cup \{A\}$
				\item $f^{cd}_{V}(x_V,e)=\left\{\begin{array}{l} f_V(x_V,e) \text{ , if } e \in \Gamma_V(x_V)\\[0.15cm] A \text{ , if  } e \in \Eoa \wedge e \notin \Gamma_V(x_V)\end{array}\right.$
				\item $f^{cd}_{V}(A,e)=A$ for all $e \in \Euc\cup\Ecva$
			\end{itemize}
			%
			\State Build $G_T=G_V^{cd}||G_F$, where $G_F$ is defined in Algorithm 1 in \citep{Moreira:2011} 
			\If {there exists $\{x^{cd}_V,(x,\ell)\}$ in $G_T$ such that $x_V^{cd}=A$ and $x \in \Xfm$} SafeControllability = false
			\Else{
				SafeControllability = true
				\EndIf}
			\EndIf}
	\end{algorithmic}
\end{algorithm}

\begin{proposition}\label{thm:ver}
	Let $L_M$ denote the language generated by $G_M$. Then, $L_M$ is not AE-safe controllable with respect to $\Po^a:\Ea^*\rightarrow\Eoa^*$, $\Ecva$ and $\Xfm$ if and only if at least one of the conditions holds true
	
	\begin{enumerate}
		\item There exists $x_V=\{(x_N,N),(x,Y)\} \in X_V$ such that $x \in \Xfm$, where $x_N \in X_N$ and $x \in X_M$.
		\item There exists $\{x^{cd}_V,(x,Y)\} \in X_T$ such that $x_V^{cd}=A$ and $x \in \Xfm$, where $x^{cd}_V \in X^{cd}_V$ and $x \in X_M$.
	\end{enumerate}
\end{proposition}

\proof Given in Appendix.

\begin{example}
	\label{ex:ae_ver}
	Returning again to example \ref{ex:ex1}, the closed-loop system subject to actuator enablement attacks is shown in Figure \ref{fig:ex1_Gm} where the sets of observable, controllable, and vulnerable actuator events are $\Eo=\Eset$, $\Ec=\{b\}$, and $\Ecv=\{b\}$, respectively.
	\begin{figure}[h]
		\center
		\subfigure[\label{fig:ex1_GN}Non-fault automaton $G_N$]{\includegraphics{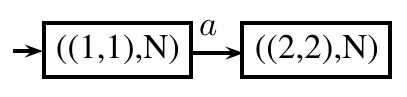}}\\
		
		\subfigure[\label{fig:ex1_GV} Verifier automaton $G_V$]{\includegraphics{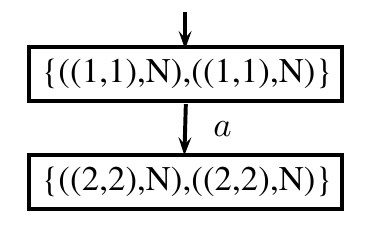}}\\
		
		\subfigure[\label{fig:ex1_GVcomp} Verifier automaton $G_V^{cd}$]{\includegraphics{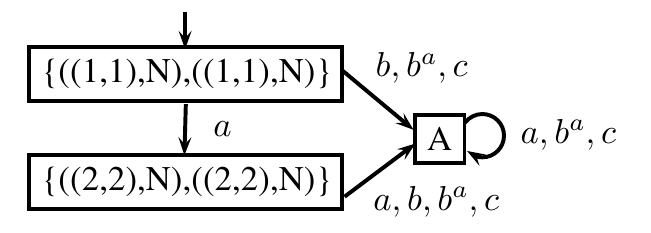}}\\
		
		\subfigure[\label{fig:ex1_GT} Automaton $G_T$]{\includegraphics{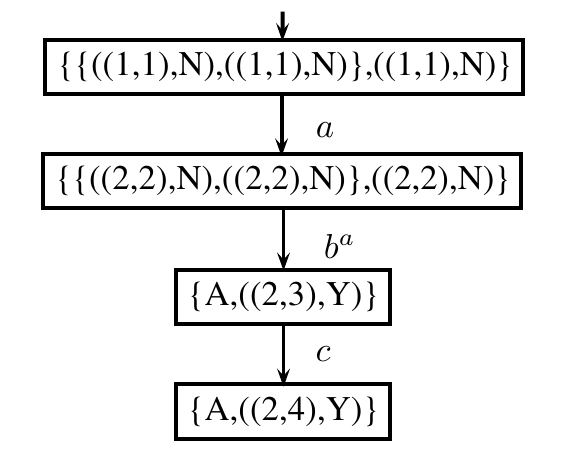}}\\

		\caption{Figures of Example \ref{ex:ae_ver}}
	\end{figure}
	The normal and the faulty/attacked behavior of the system $G_N$ and $G_F$ are depicted in Figures \ref{fig:ex1_GN} and \ref{fig:ex1_Gl}, respectively, and verifier $G_V$ is shown in Figure \ref{fig:ex1_GV}.
	According to Step 5 of Algorithm \ref{alg:scv}, it is necessary to add a new state $A$.
	All states of $G_V$ are connected to $A$ using observable events $a,b,b^a$ and $c$ (when these events are in the active event set of the state). Also, it is necessary to add self-loops at state $A$ for uncontrollable events $a,b^a$ and $c$, as shown in Figure~\ref{fig:ex1_GVcomp}.
	After that, $G_T$ is built by computing $G_V^{cd}\|G_F$ as depicted in Figure \ref{fig:ex1_GT}.
	The system is not AE-safe controllable according to step 7 of Algorithm \ref{alg:scv}, because state $\{A,((2,4),Y)\}$ in $G_T$ has, as components, state $A$ and  $(2,4) \in \Xfm$. Thus, the supervisor cannot prevent the system from reaching an unsafe state after the system is sure that an attack has occurred.
\end{example}

}
\subsection{Discussion} 
Recall from Algorithm~\ref{alg:ae} that we model  AE-attacks by adding in the supervisor realization $H$ a self-loop for every $\sigma^a \in \Ecva$ (unless the compression of $\sigma^a$ is already in the active event set of the state).
The resulting automaton $H_a$  thus models an  ``all-out"  attacker that always attacks the vulnerable actuators.
Subsequently, AE-safe controllability is a property of whether the system can be protected under such an all-out  attacker.
Now, we consider the question of whether it is possible under ``smaller'' attacks, i.e., when the attacker does not attack at all times, to inflict  damage on the system when AE-safe controllable holds.
The following proposition proves that AE-safe controllability with respect to the all-out attacker implies AE-safe controllability with respect to any attacker.
Hence, testing AE-safe controllability with respect to the all-out attacker is sufficient.
\begin{proposition}
	\label{thm:mattacks}
	Let $L_{AA}$ be the language of the closed-loop system under the all-out attacker and $L_{SA}$ be that under an attacker that does not attack at all times. If $L_{AA}$ is AE-safe controllable with respect to $\Po$, $\Ecva$ and $X_f^M$, then $L_{SA}$ is AE-safe controllable with respect to $\Po$, $\Ecva$ and $X_f^M$.
\end{proposition}

\proof Given in Appendix.

\subsection{Actuator disablement attacks} 

We briefly discuss actuator disablement attacks (AD-attacks), which correspond to the case where the fusion block $+/-$ in Figure~\ref{fig:attacks} is the conjunction of the enabled events of $S_P$ with those of $A$; that is, vulnerable actuator events that are enabled by the supervisor can be disabled by the attacker.
In this case, the closed-loop behavior is further restricted to a subset of $K$, since no new behavior of $G$ can be generated.
Hence, no state in $X_f$ is reachable.
However, blocking may occur, even if the closed-loop system $S_P/G$ is nonblocking; an example can be easily constructed and is omitted here.
Clearly, the only motivation for $A$ to select such an attack is to cause blocking.
We will not further discuss this type of attack since it cannot lead to a violation of safety, as described by $X_f$.

\section{Sensor erasure attacks} 
	\label{sec:se}

In this and the next two sections, we discuss attacks on vulnerable sensors.
We first consider the case of sensor erasure attacks, or SE-attacks.
As illustrated in Figure~\ref{fig:attacks}, in SE-attacks, the attacker $A$ can ``erase'' an occurrence of an observable event $\ev \in \Eov$ to $S_P$ and $G_d$. 
Thus, enabled observable events in $\Eov$ can be ``subsumed'' by corresponding \emph{unobservable}  events that we label as the set $\Eova$, thereby causing confusion for both $S_P$ and $G_D$.
%
%
Hence, using $\Eova$,
the modeling of SE-attacks follows a similar procedure as for AE-attacks in Algorithm~\ref{alg:ae}, with some minor adjustments.
For the sake of clarity, the modified form of Algorithm~\ref{alg:ae} is given in Algorithm~\ref{alg:se}.
%
Note that the events in $\Eova$ are necessarily unobservable and that their controllability properties 
are inherited from the corresponding ones in $\Eov$.

	
	\begin{algorithm}[htb]
		\caption{Algorithm for SE-attack model}
		\label{alg:se}
		\begin{algorithmic}[1]
			\Statex \textbf{Inputs:}\begin{itemize} \item $G=(X,E,f,x_0)$: plant
				\item $H=(X_{H},\Eset,f_{H},x_{0,{H}})$ : supervisor realization 
				\item $\Eo$, $\Ec$ and $\Eov$ : sets of observable, controllable, and vulnerable sensor events 
				\end{itemize}
			\Statex \textbf{Output:} 
				\begin{itemize}  \item $G_M=(X_M,\Ea,f_M,x_{0,m})$: closed-loop system subject to SE-attacks								\end{itemize}
			\State Compute $\Eova=D(\Eov)\backslash\Eov$
			\State Define $\Ea=E \cup \Eova$ and assign:
			\begin{itemize}
				\item $\Eoa=\Eo$ 
				\item $\Euoa=\Euo\cup\Eova$
				\item $\Eca=\Ec\cup D(\Eov\cap\Ec)$
				\item $\Euca=\Euc\cup D(\Eov\cap\Euc)$
			\end{itemize}	
			\State Build $G_a=(X,E_{a},f_{a},x_{0})$, where
			\begin{itemize}
				\item $f_a(x,\ev^a)=f(x,C(\ev^a))$ if $f(x,C(\ev^a))$ is {defined}, $\forall \ev^a \in \Ea $, $\forall x \in X$
			\end{itemize}
			\State Build $H_a=(X_{H},E_a,f_{H_a},x_{0,{H}})$, where
			\begin{itemize}
				\item $f_{H_a}(x_H,\ev) = $
				
				$\begin{cases}  f_H(x_H,\ev), & \text{ if } f_H(x_H,\ev) \text{ is {defined}}, \\ x_H,  & \text{ if } (\ev \in \Eova \wedge \\ & f_H(x_H,C(\ev)) \text{ is defined}) \\ & \vee (\ev \in \Euca \wedge f_{H}(x_H,\ev) \\ &\text{ is undefined})\end{cases}$
				
				
			\end{itemize}
			\State Compute $G_M=H_a\|G_a$
		\end{algorithmic}
	\end{algorithm}
To explain the reasoning behind step 3 of the algorithm, we make the following observations.
The erasure of (enabled) observable events means that the supervisor and $G$ may become ``out of sync'' from the original design of $H$; this is why all uncontrollable events in $\Euc$ must be added at all states of $H$, if they are not already there, to make sure that controllability is never violated.
The same reasoning applies to all events in $\Eova$ that are uncontrollable, as the occurrence of an uncontrollable vulnerable event could be erased by $A$.
However, a \emph{controllable} event in $\Eov$ will only be erased, i.e., replaced by its corresponding event in $\Eova$, if it is enabled by $H_a$.
In all cases, feasibility in $G$ of the self-loops added in $H_a$ will be captured by the parallel composition $H_a\|G_a$.
In this manner, the construction of $G_M$ again captures the case where $A$ may attack at every possible opportunity, i.e., it may erase every event output by a vulnerable sensor.

	\begin{example}
		\label{ex:SE}
		The state transition diagrams of system $G$ and supervisor realization $H$ are shown in Figures \ref{fig:SEexampleG} and \ref{fig:SEexampleH}, respectively, where $\Ec=\{a,c\}$ and $E = \Eo=\{a,b,c\}$. 
		\begin{figure}[h]
				\center
				\subfigure[\label{fig:SEexampleG}  $G$]{\includegraphics[width=0.25\textwidth]{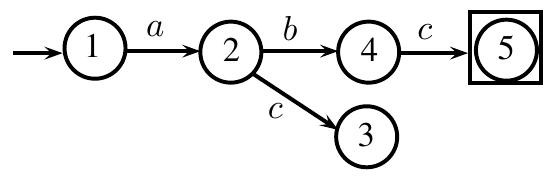}}\\
				
				\subfigure[\label{fig:SEexampleH}  $H$]{\includegraphics[width=0.2\textwidth]{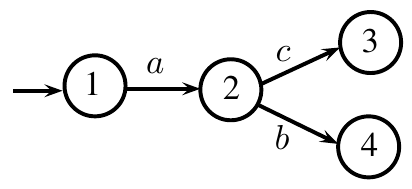}}\\
				
				\subfigure[\label{fig:SEexampleGa}  $G_a$]{\includegraphics[width=0.25\textwidth]{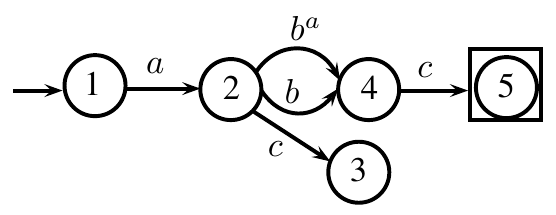}}\\				
				
				\subfigure[\label{fig:SEexampleHA}  $H_a$]{\includegraphics[width=0.2\textwidth]{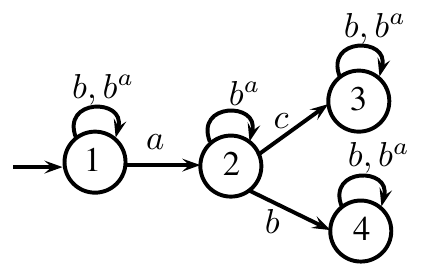}}\\
				
				\subfigure[\label{fig:SEexampleGM}  $G_M$]{\includegraphics[width=0.3\textwidth]{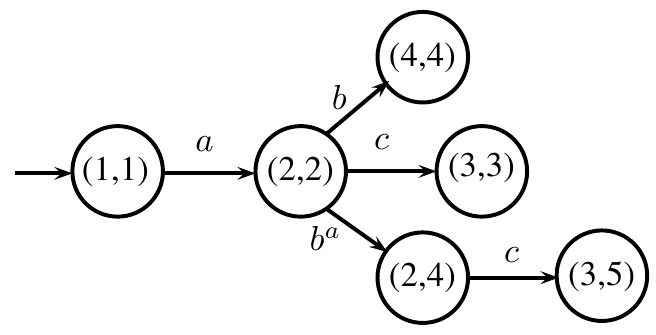}}
								
				\caption{Figures of Example \ref{ex:SE}}
		\end{figure}
		Let $\Eov=\{b\}$ be the set of vulnerable (to erasure) sensor events. The set of unsafe states is $X_f=\{5\}$, marked with a square in Figure \ref{fig:SEexampleG}. 
		The erased sensor event set is $\Eova=\{b^a\}$, thus the new set of events is $\Ea=\{a,b,c,b^a\}$. 
		Following Algorithm~\ref{alg:se}, we build automaton $G_a$ by adding a transition labeled by $b^a$ in parallel with every transition $b$, as shown in Figure \ref{fig:SEexampleGa}. 
		The realization of the supervisor under SE-attacks is depicted in Figure \ref{fig:SEexampleHA}.
		The closed-loop system under SE-attacks is computed as $G_M=H_a\|G_a$, and it is shown in Figure~\ref{fig:SEexampleGM}. After the occurrence of event $a$, the attacker erases the occurrence of event $b$ and it becomes unobservable to the supervisor. Because of that, the supervisor ``thinks'' that the plant is in state $2$, but the plant is actually in state $4$. 
		Then, the supervisor allows event $c$ to occur and the plant reaches an unsafe state.
		
	\end{example}

The next example shows that SE-attacks can lead to  blocking, when marked states are considered in $G$.
	\begin{example} 
		\label{ex:BlockingSE}
		Consider $G$ and $H$
		 in Figures \ref{fig:BSEexampleG} and \ref{fig:BSEexampleH}, respectively, where $E = \Eo = \Ec$. 
		\begin{figure}[h]
				\center
				\subfigure[\label{fig:BSEexampleG}  $G$]{\includegraphics[width=0.25\textwidth]{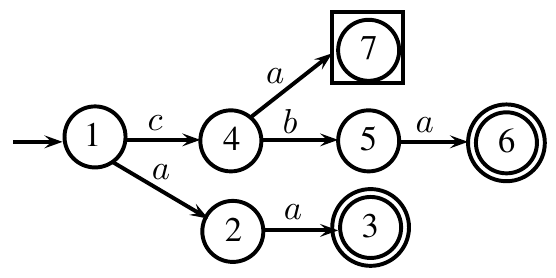}}\\
				
				\subfigure[\label{fig:BSEexampleH}  $H$]{\includegraphics[width=0.25\textwidth]{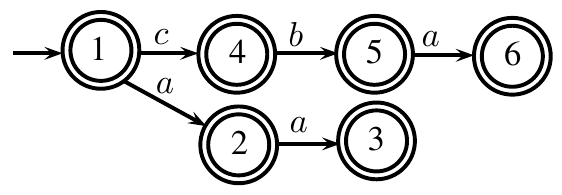}}\\
				
				\subfigure[\label{fig:BSEexampleGM}  $G_M$]{\includegraphics[width=0.3\textwidth]{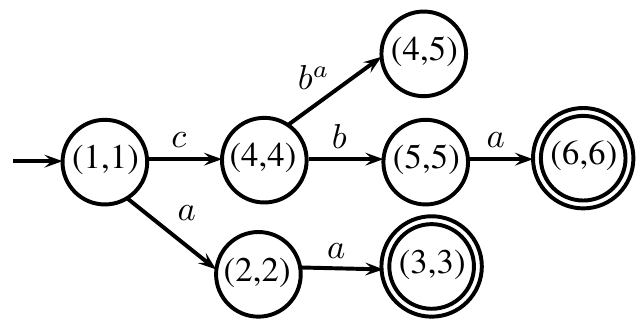}}	
							
				\caption{Figures of Example \ref{ex:BlockingSE}}
		\end{figure}
		Let $\Eov=\{b\}$ and let $X_f=\{7\}$.
				Following Algorithm~\ref{alg:se}, the closed-loop system subject to SE-attacks is shown in Figure~\ref{fig:BSEexampleGM}. 
When the attacker erases event $b$, the plant gets stuck in state 5 as the supervisor assumes the plant is in state 4.
%
		
	\end{example}

\section{Sensor insertion attacks} 
	\label{sec:si}
	Sensor insertion attacks, or SI-attacks for short, can ``insert'' a fictitious occurrence of an observable event $\ev \in \Eo$ to the observation stream of supervisor $S_P$ and intrusion detection module $G_D$.
In order to model SI-attacks, let $\Eovi=\{\ev^i : \ev \in \Eov\}$ denote the set of attacks by $A$ on the vulnerable sensors, which we will refer to as \emph{SI-attack onset events}. 
%

The case of SI-attacks is somewhat different from the attacks previously considered in this paper, in that we need to be more specific about the attack strategy of $A$.
Namely, if $A$ inserts a fictitious event occurrence that is not defined at the current state of $H$, either because the event is not currently feasible in the state $H$ thinks the system is in or because it is currently disabled by $S_P$, then $A$ immediately reveals its presence without gaining any benefit.
Hence, it is only advantageous for $A$ to insert fictitious event occurrences when $H$ expects such observations.
The goal of $A$ in this case is to cause a change of control action for $S_P$ that would, for instance, enable an event that was not currently enabled in order to steer $G$ towards an unsafe state.
To resolve the above issue, we will assume that in the case of SI-attacks, $A$ has a model of $H$ and moreover $A$ has the same observation capabilities as $S_P$; hence, $A$ knows at any time the exact state of $S_P$.
Under this assumption, $A$ will only insert fictitious event occurrences when $H$ expects that such an event could have occurred.

The modeling of the closed-loop system under the above considerations is obtained by executing Algorithm~\ref{alg:si}. 
Consider plant $G$ and supervisor realization $H$. 
First, we construct $G_a$ by creating a new state $x_{\ev}^{j}$for each event $\ev \in \Eov$ at every state $j$ of $G$, and a new event $\ev^i \in \Eovi$ {that represents the onset of an SI-attack} at that state of $G$.
Next, for each added state, we add two transitions: one from state $j$ to the new state $x_{\ev}^{j}$ labeled by $\ev^i \in \Eovi$, and the other from the new state $x_{\ev}^{j}$ to state $j$ labeled by $\ev$; the former represents the onset of an SI-attack on the vulnerable sensor of the plant whereas the latter represents the fictitious event inserted by the SI-attack.
This inserted fictitious event is assumed to be indistinguishable from a genuine one by $S_P$ and $G_D$, which is why the latter transition is labeled by $\ev \in \Eo$.

%

Afterwards, we build automaton $H_a$ that models the realization of the supervisor under SI-attack. 
For this purpose, we add a self-loop for (unobservable) event $x_{\ev}$ at each state $x$ that has event $\ev \in \Eov$ in its active event set.
This ensures that the attacker can insert any occurrence of any event in $\Eov$ when such an event is feasible according to the original design of $H$, which is consistent with the above-described attack model.
Moreover, to ensure controllability, we also add at each state $x$ a self-loop for every uncontrollable event in $\Euc$ that is not already in the active event set of state $x$.
Finally, the closed-loop system subject to SI-attacks is obtained by the parallel composition of $H_a$ and $G_a$.

	\begin{algorithm}[htb]
		\caption{Algorithm for SI-attack model}
		\label{alg:si}
		\begin{algorithmic}[1]
			\Statex \textbf{Inputs:}\begin{itemize} \item $G=(X,E,f,x_0)$ and $H=(X_{H},\Eset,f_{H},x_{0,{H}})$ : plant and supervisor realizations, respectively \item $\Eo$, $\Ec$, and $\Eov$ : sets of observable, controllable, and vulnerable sensor events \end{itemize}
			\Statex \textbf{Output:} \begin{itemize}  \item $G_M=(X_M,E_M,f_M,x_{0,m})$: closed-loop system subject to SI-attacks\end{itemize}
			\State $G_a \gets$\Call{Build-Ga}{G} 
			\State $H_a \gets$\Call{Build-Ha}{H} 
			\State Compute $G_M=H_a\|G_a$
			
			\Function{Build-Ga}{$G$}
			\State $G_a\gets G$
			\State $\Ea\gets E \cup \Eovi$
			\For{ every $j \in X$} 
			\For{ every $\ev \in \Eov$}
			\State $X_a \gets X_a \cup \{ x_{\ev}^j\}$			
		        \State Add $f_a(j,\ev^i)=x_{\ev}^j$
				\State Add $f_a(x_{\ev}^j,\ev)=j$
			\EndFor
			\EndFor	
			\Return{$G_a=(X_a,E_{a},f_{a},x_{0})$}	
			\EndFunction

			\Function{Build-Ha}{$H$}
			\State $H_a \gets H$	
			\State $E_{H_a}   \gets \Eset \cup \Eovi$
			\For{all $x \in X_H$}
			\State Add $f_{H_a}(x, \ev^i)=x$ for all $\ev \in (\Gamma_H(x)\cap \Eov)$
			\State Add $f_{H_a} (x, \ev) = x$ for all $\ev \in \Euc \setminus \Gamma_H(x)$  
			\EndFor	
			\Return{$H_a=(X_{H_a},E_{H_a},f_{H_a},x_{0,{H}})$}
			\EndFunction
			
		\end{algorithmic}
	\end{algorithm}

	\begin{remark}
		The events in $\Eovi$ are unobservable and uncontrollable, as they represent the onset of an SI-attack.
		Hence, we have that:
		(i) $\EoM=\Eo$;
		(ii) $\EuoM=\Euo\cup\Eov^i$;
		and
		(iii) $\EucM=\Euc \cup\Eov^i$.
	\end{remark}

	\begin{example}
		\label{ex:SI}
		Consider system $G$ and of supervisor realization $H$ shown in Figures \ref{fig:SIexampleG} and \ref{fig:SIexampleH}, respectively, where $E = \Eo = \Ec$. 
		\begin{figure}[h]
			\center
			\subfigure[\label{fig:SIexampleG}  $G$]{\includegraphics[width=0.25\textwidth]{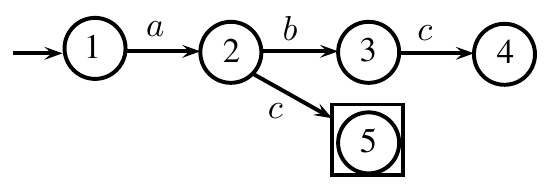}}\\
			
			\subfigure[\label{fig:SIexampleH}  $H$]{\includegraphics[width=0.25\textwidth]{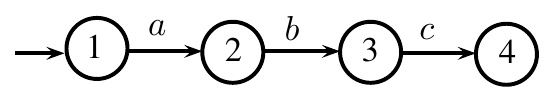}}\\
			
			\subfigure[\label{fig:SIexampleGa}  $G_a$]{\includegraphics[width=0.25\textwidth]{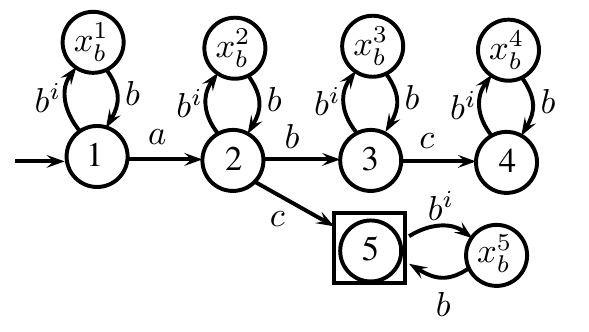}}\\				
			
			\subfigure[\label{fig:SIexampleHA}  {$H_a$}]{\includegraphics[width=0.25\textwidth]{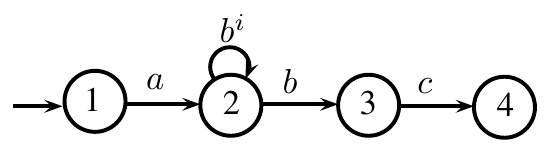}}\\
			
			\subfigure[\label{fig:SIexampleGM}  $G_M$]{\includegraphics[width=0.3\textwidth]{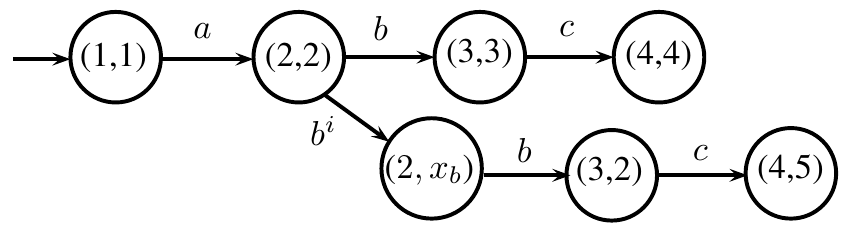}}
			
			\caption{Figures of Example \ref{ex:SI}}
		\end{figure}
		Let $\Eov=\{b\}$ and $X_f=\{5\}$. 
		
		Following Algorithm \ref{alg:si}, automaton $G_a$ is built by adding new states  and transitions labeled by  $b^i$  and $b$  as shown in Figure \ref{fig:SIexampleGa}. 
		The idea behind this procedure is that when the SI-attack occurs, which is represented by event $b^i$, the attacker emulates the occurrence of event $b$ in the plant $G$; hence $G_a$ does not reach a new state.
		The new set of events is $\Ea=\{a,b,c,b^i\}$, where $\Euoa=\{b^i\}$. 
		The realization of the supervisor subject to  SI-attacks is shown in Figure \ref{fig:SIexampleHA}.
		Supervisor $H_a$ does not see the difference between the fictitious inserted event $b$ (after $b^i$) and the real occurrence of $b$ (with no $b^i$), hence it changes state in both cases. 
		The closed-loop system under SI-attacks is computed as $G_M=H_a\|G_a$ and it is shown in Figure \ref{fig:SIexampleGM}. After the onset attack event $b^i$ followed by inserted event $b$, the supervisor ``thinks'' that the system is in state $3$; however, the system is in state $2$. Then, the supervisor enables event $c$ and the plant can reach an unsafe state. 
		
	\end{example}

\section{General approach for detection of attacks} 
\label{sec:GF}

Our strategy for detection and mitigation of sensor attacks is the same as that for AE-attacks, described in Section \ref{sec:det_mit_attacks}.
The intrusion detection module monitors the output from the plant and notifies the supervisor when an attack has been detected.
The supervisor, upon receiving an attack report from the intrusion detection module, switches to a {safe mode} of operation where it disables all controllable events. 
Hereafter, we generalize the property of AE-safe controllability to a general form that captures, in a unified manner, AE-, SE-, and SI-attacks.
Then we discuss its verification.

\subsection{General form of safe controllability}
We defined in Section \ref{sec:AE-SC} AE-safe controllability, which ensures that AE-attacks can be detected in time to  avoid reaching an unsafe state, under the attack and defense strategies considered in this paper.
We now generalize AE-safe controllability to a General Form termed \emph{GF-Safe Controllability}.
To avoid ambiguity, we denoted the event set of $G_M$ as $\EM$ and specify it in each case.

\begin{definition}[GF-Safe Controllability]\label{def:GFsafcon}
	Consider $G_{M}=(X_M,\EM,f_M,x_{0,M})$ a model of one type of attacks (AE, SE, or SI). 
	Language $L_M= {\cal L}(G_{M})$  is GF-safe controllable with respect to projection $\Po^M : \EM^*\rightarrow \Eo^*$,  set of vulnerable events $\Ef$, and unsafe states $X_f^M$ if	
	$(\forall s \in \Psi(\Ef))(\forall t \in L_M/s,\: t \in \EM^*)$ $[(f_M(x_o,st) \cap X_f^M \neq \emptyset) \wedge (\forall s'<  st, f_M(x_o,s') \cap X_f^M = \emptyset)] \Rightarrow$
	$ (\exists t_1,t_2 \in \EM^*) [(t = t_1 t_2) \wedge \left((\nexists \omega \in L_M)[\Po^M(st_1)=\Po^M(\omega) \wedge \Eova \notin \omega]\right)\wedge (\Ec \in t_2)]$.
\end{definition}

Comparing Definitions~\ref{def:safcon} and~\ref{def:GFsafcon}, we see that by replacing $\Po^M$  and $\Ef$ by $\Po^a$ and $\Ecva$, respectively, then GF-safe controllability reduces to AE-safe controllability.
GF-safe controllability allows us to address SE-attacks and SI-attacks as well. 
For SE-attacks, $G_M$ is obtained using Algorithm~\ref{alg:se}, as shown in Section~\ref{sec:se}, where $\EM = \Eset \cup \Eova$, projection $\Po^M$ is $\Po^M:(\Eset \cup \Eova)^*\rightarrow\EoM^*$, and $\Ef$ is the set of erased sensor events $\Eova$. 
For SE-attacks, $G_M$ is obtained using  Algorithm~\ref{alg:si}, where $\EM = \Eset \cup \Eovi$, projection $\Po^M$ is $\Po^M:(\Eset \cup \Eovi )^*\rightarrow \EoM^*$, and $\Ef$ is the set of onset attack events $\Eovi$.
We will refer to GF-safe controllability for SE- and SI-attacks as \emph{SE-safe controllability} and \emph{SI-safe controllability}, respectively. 

Clearly, Theorem~\ref{thm:one} generalizes to the case of SE- and SI-attacks, using SE-safe controllability and SI-safe controllability, respectively, since once the modified model $G_M$ that accounts for attacks has been built, then the conditions for avoiding unsafe states boil down to the same cases in each attack type.

\subsection{Test of GF-Safe Controllability}\label{sec:testGM} 
To test if a system is GF-safe controllable, we generalize Algorithm~\ref{alg:sc} to Algorithm~\ref{alg:ss}.
This algorithm verifies if the intrusion detection module can detect any attack before the plant reaches an unsafe state and if the supervisor can disable events to prevent the plant from reaching $X_f$. 
Here, the label automaton to use for building the diagnoser is parametrized by $\Ef$, as shown in Figure~\ref{fig:label-GFsc}.
%
%
\begin{figure}[htb]
	\centering
	\includegraphics[width=0.15\textwidth]{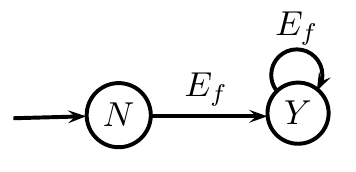}
	\caption{Label automaton $A_\ell^{GF}$}
\label{fig:label-GFsc}
\end{figure}
In step 1 of  Algorithm~\ref{alg:ss}, we select which attack we want to analyze. 
After that, we follow the same steps as in Algorithm~\ref{alg:sc}. 
For the sake of brevity, we omit explaining each step.
We note that in the algorithm, $q_{\downarrow x} := \{x: (\exists l)[(x, l) \in q]\}$ is the projection of $q$ to the set of corresponding $G_M$ states.
\begin{algorithm}[htb]
	\caption{GF-safe controllability test}
	\label{alg:ss}
	\begin{algorithmic}[1]
		\Statex \textbf{Inputs:}\begin{itemize} \item $G_{M}=(X_M,\EM,f_M,x_{0,M})$: closed-loop system subject to AE-attacks, SE-attacks or SI-attacks
			\item $\Xf^M$: set of unsafe states in $G_M$
			\item $\Ef \in \{\Ecva,\Eova,\Eovi \}$
			\item AttackMode $\in \{AE,SE,SI\}$
			\end{itemize}
		\Statex \textbf{Output:} GFSafeControllability $\in \{true,false\}$			
			\If{AttackMode=AE} 
			\State $\Ef=\Ecva$ 
			\Else \If {AttackMode=SE}
			\State $\Ef=\Eova$
			\Else \State $\Ef=\Eovi$ 
			\EndIf
			\EndIf
		\State Build $G_\ell= G_{M} || A_\ell^{GF}$ where $A_\ell^{GF}$ is shown in Figure \ref{fig:label-GFsc}
		\State Compute diagnoser $G_d=Obs(G_\ell,\EuoM )$, where $\EuoM=\Euo\cup D(\Ecv\cap\Euo)$ if $AttackMode$ is $AE$, otherwise $\EuoM=\Euo\cup\Ef$
		\State Verify for all uncertain states $q_{YN}=\{(x_{i_1},\ell_{i_1}), \ldots (x_{i_n},\ell_{i_n})\}\in X_d $, where $x_{i_k} \in X_H\times X$ and $\ell_{i_k} \in\{Y,N\}$, $k \in \{1,\ldots, n\}$ if $\exists j \in \{1,\ldots ,n\}$ such that $x_{i_j} \in \Xfm$
			\If {there is uncertain state $q=\{(x_{i_1},\ell_{i_1}),\ldots, (x_{i_n},\ell_{i_n})\}$\Statex{$\in Q_{YN} $ in which there exists $x_{i_j} \in \Xfm$}}
			\State{GFSafeControllability $=$ false}
			\Else{
				Compute $\cal{FC}$ according to Definition \ref{def:FCS}
				\If {there is $q=\{(x_{i_1},Y), \ldots, (x_{i_n},Y)\}\in \mathcal{FC} $  in which there exists $x_{i_j} \in \Xfm$} 
				\State GFSafeControllability $=$ false
				\Else{
					\State Compute \small{$$X^{uc}=\bigcup_{q \in \mathcal{FC}} \bigcup_{x_{M}\in q_{\downarrow x}} Reach(G_{M},x_{M},\EucM),$$ where 
						\If {AttackMode=AE} \State $\EucM=\Euc\cup\Ecva$
						\Else \If {AttackMode=SE} \State $\EucM=\Euc\cup D(\Eov\cap\Euc)$ 
						\Else \State $\EucM=\Euc\cup\Eovi$ 
						\EndIf\EndIf}} 
					         
					\If {$ X^{uc} \cap \Xfm \neq \emptyset$ } 
					\State GFSafeControllability $=$ false
					\Else { GFSafeControllability $=$ true}
					\EndIf
				\EndIf}
			\EndIf	
	\end{algorithmic}
\end{algorithm}


\begin{proposition}\label{thm:diagG}
	Let $G_{M}=(X_M,\Ea,f_M,x_{0,M})$ be obtained from one of Algorithms~\ref{alg:ae}, \ref{alg:se}, \ref{alg:si} presented earlier,
	 and let 
	automaton $G_d$ be the diagnoser built in Algorithm \ref{alg:ss}. 
	Language $L_M$ is \emph{not} GF-safe controllable with respect to $\Po^M$, $\Ef$, and $\Xfm$ if and only if one of the following conditions holds true:
	\begin{enumerate}
		\item There exists $q_{YN}=\{(x_{i_1},\ell_{i_1}), \ldots, (x_{i_n},\ell_{i_n})\}\in Q_{YN} $ 
		such that $\exists j \in \{1,\ldots, n\}$, $x_{i_j} \in \Xfm$ and $\ell_{i_j}=Y$.
		\item There exists $q_{Y}=\{(x_{i_1},Y), \ldots, (x_{i_n},Y)\}\in \mathcal{FC} $ 
		such that $\exists j \in \{1,\ldots ,n\}$, $x_{i_j} \in \Xfm$.
		\item There exists $x_M \in X^{uc}$ such that $x_M \in \Xfm$, where $X^{uc}$ is defined in Algorithm \ref{alg:ss}.
	\end{enumerate}
\end{proposition}

\proof The proof follows the same steps as the proof of  {Proposition~\ref{thm:diag}}.

\begin{remark}
	A test of GF-safe controllability using verifiers can be obtained in a straightforward manner by suitably adapting Algorithm~\ref{alg:scv} to the different attack cases. We omit the details.
\end{remark}
\begin{example}
	Let us consider again Example \ref{ex:SE}, whose diagnoser $G_d$ built according to Algorithm \ref{alg:ss} is drawn in Figure \ref{fig:SEexampleGd} considering $\Eova=\{b^a\}$. 
	By examining the diagnoser states, we see that the attack guides the system to unsafe state $\Xf=\{5\}$, and the state $\{((3,3)N),((3,5)Y)\}$ is an uncertain state. Thus, the system under SE-attack represented in Figure \ref{fig:SEexampleG} is not SE-safe controllable with respect to $\Po^M$, $\Eova$ and $\Xf$.
	\begin{figure}[htb]
		\centering
		\includegraphics[width=0.2\textwidth]{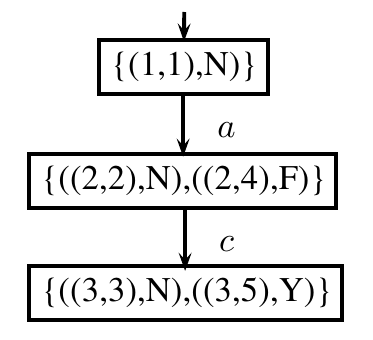}
		\caption{\label{fig:SEexampleGd} Automaton $G_d$ for Example \ref{ex:SE}:  {\it SE-attack}}
	\end{figure}
	
	Consider now Example \ref{ex:SI}, where the closed-loop system under SI-attacks is shown in Figure \ref{fig:SIexampleGM}. We follow Algorithm~\ref{alg:ss} to test if the language generated by this $G_M$ 
	is SI-safe controllable. The diagnoser $G_d$ for this case is drawn in Figure \ref{fig:SIexampleGd}. By examining the diagnoser states, we notice that the attack will not be detected by the diagnoser and the system will reach unsafe state $\Xf=\{5\}$. Thus, the system is not SI-safe controllable with respect to $\Po^M$, $\Eovi$ and $\Xf$.
	\begin{figure}[htb]
		\centering
		\includegraphics[width=0.25\textwidth]{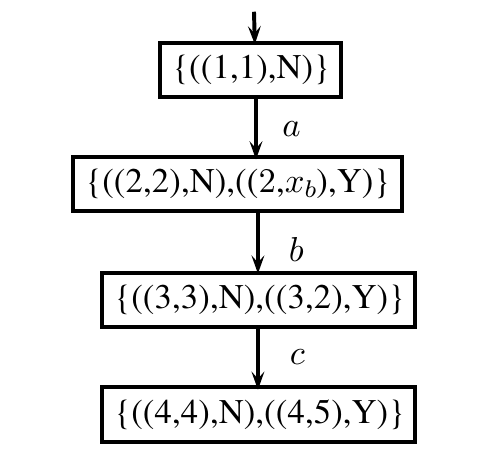}
		\caption{\label{fig:SIexampleGd} Automaton $G_d$ for Example \ref{ex:SI}: {\it SI-attack}}
	\end{figure}
	
\end{example}

\begin{remark}
After testing GF-safe controllability, it may also be necessary to test if the new closed-loop system has a deadlock in the case of SE- or SI-attacks; recall Example~\ref{ex:BlockingSE} (a similar example can be constructed for SI-attacks). 
\end{remark}

\subsection{Combinations of attacks}

So far, for the sake of simplicity of presentation, we have considered attacks of a single type.
However, our methodology is general and combinations of attacks can be studied by overlapping the modeling approaches of each attack in the construction of $G_M$, via $G_a$ and $H_a$. As illustration, we sketch a modification to Example 4 which requires a combination attack to bring the system to an unsafe state. Make event $b$ controllable. Add a new state $6$ and a controllable and observable event $d$ that takes state $3$ to $6$. Also replace event $c$ that takes state $4$ to $5$ with event 
$d$. Now erasure of $b$ alone will not produce a successful attack, but the combination of sensor erasure of $b$ and insertion of $c$ at state $2$ will be successful.
The details are straightforward and omitted here.
%
%
Notice that, for the combination of AE- and SE- attacks, we need to adjust the superscripts of the two dilations for AE- and SE- attacks since we used  the same superscript ``a'' for each, but with a different meaning in each individual case. 
In AE-attacks, the event that was created by the dilation is uncontrollable whereas, in SE-attacks, it is unobservable. Therefore, when considering a combination of these two attacks, one should use two different superscripts.

In addition, the construction of $G_M$ could also be altered to model attackers that do not attack at every opportunity, assuming some knowledge of the attack model in terms of the state spaces of $G$ and $H$.
How to acquire such knowledge is an interesting problem for future research, but one that is likely to be application-dependent.


\section{Traffic control system example} 
\label{sec:exemple}

As a more comprehensive example than the illustrative ones presented so far,
let us consider an attack on a  traffic control system taken from \cite{Wonham:2013} and often used in the literature. 
The problem consists of two vehicles $a$ and $b$ that must travel from the origin to the destination through a single one-way road. The road is partitioned into four sections.
Denote by $a_i$ the events corresponding to vehicle $a$ entering section $Si$, $i=1,\ldots,5$, where $S5$ is the destination. Similarly, we have events $b_i$ for vehicle $b$.
The plant behavior is modeled by $G=G_{va}||G_{vb}$, where $G_{va}$ and $G_{vb}$ are such that ${\cal L}(G_{va})=\overline{\{a_1a_2a_3a_4a_5\}}$ and ${\cal L}(G_{vb})=\overline{\{b_1b_2b_3b_4b_5\}}$.
Traffic lights and vehicle detectors are installed at section junctions.
The traffic lights are placed at the entrances of sections 1, 2, and 4, and the vehicle detectors are located at the entrances of sections 1, 3, 4, and at the destination. 
Thus, $\Ec=\{a_1,b_1,a_2,b_2,a_4,b_4\}$ and $\Eo=\{a_1,b_1,a_3,b_3,a_4,b_4,a_5,b_5\}$.
The goal is that the two vehicles must reach the destination without colliding, which is achieved by preventing both vehicles from occupying the same road section simultaneously. The unsafe states are thus the states where vehicles $a$ and $b$ are in the same section; i.e., $X_f=\{(i,i)\}$, {$i \in \{1,2,3,4\}$}. For this example, we computed a supervisor $S_P$ that satisfies this safety specification. The realization of $S_P$ was obtained as follows:
(i) states (2,1) and (1,2) were deleted to satisfy the control specification and the property of observability; (ii) the supremal controllable sublanguage was then computed; (iii) the resulting language was verified to be observable; (iv) the realization of $S_P$ was obtained in the standard manner (pp.186, \citet{cassandras2008}). This procedure was chosen as it results in a closed-loop language that is strictly larger than the supremal controllable normal sublanguage of the specification language.
	


In \cite{Carvalho:2016}, we presented an AE-attack on this traffic control system. A summary of those results is provided in the Appendix for completeness. Herein, we present SE- and SI-attacks.

Let the set of vulnerable sensor events to SE-attacks be $\Eov=\{a_3,b_3\}\subseteq\Eo$. 
Therefore, the set of attacked sensor events is $\Eova=\{a_3^a,b_3^a\}$. 
Following Algortihm \ref{alg:se}, we obtain the closed-loop system $G_M$ under attacks depicted in Figure~\ref{fig:guideway_GmSE}.
\begin{figure}[htb]
	\centering
	\includegraphics[width=0.45\textwidth]{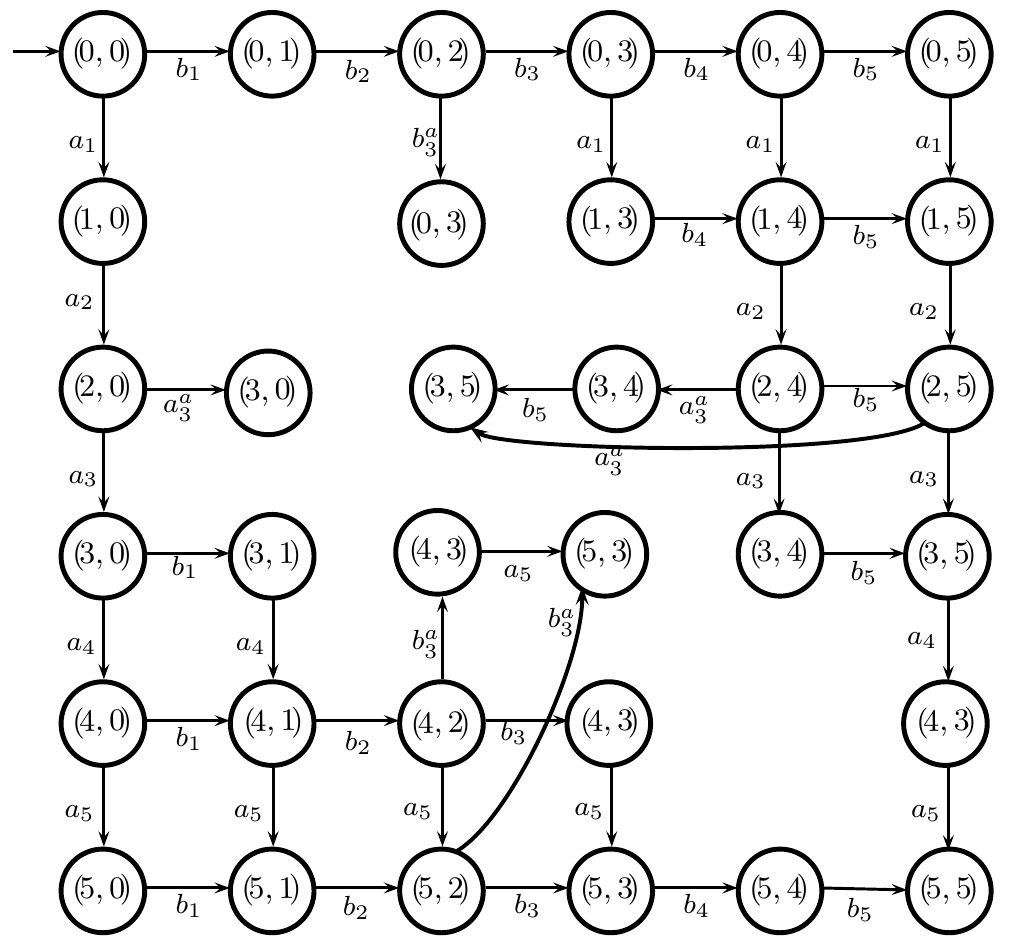}
	\caption{\label{fig:guideway_GmSE} The closed-loop system $G_M$ under SE-attack. The state labels only show the state of plant $G$.}
\end{figure}
We can see that, after attack, the system does not reach unsafe states, but deadlocks at one of the following states: $(0,3)$, $(3,0)$, $(5,3)$, and $(3,5)$. Next, we test if the system is SE-safe controllable by following Algorithm~\ref{alg:ss}. It can be seen that the system is SE-safe controllable in this case since no unsafe states are reachable in $G_M$ and, thus $ X^{uc} \cap \Xfm = \emptyset$.

Let us now consider an SI-attack where the set of vulnerable sensor events is $\Eov=\{a_4,b_4\}\subseteq\Eo$.
Thus, 
$\Eovi=\{a_4^i,b_4^i\}$. Following Algorithm~\ref{alg:si}, we obtain the closed-loop system $G_M$ under attack depicted in Figure~\ref{fig:guideway_GmSI}, where it can be seen that the system has reaches unsafe state $(3,3)$. In addition, it can be seen that the intrusion detection 

\begin{figure}[hbt]
	\centering
	\includegraphics[width=0.45\textwidth]{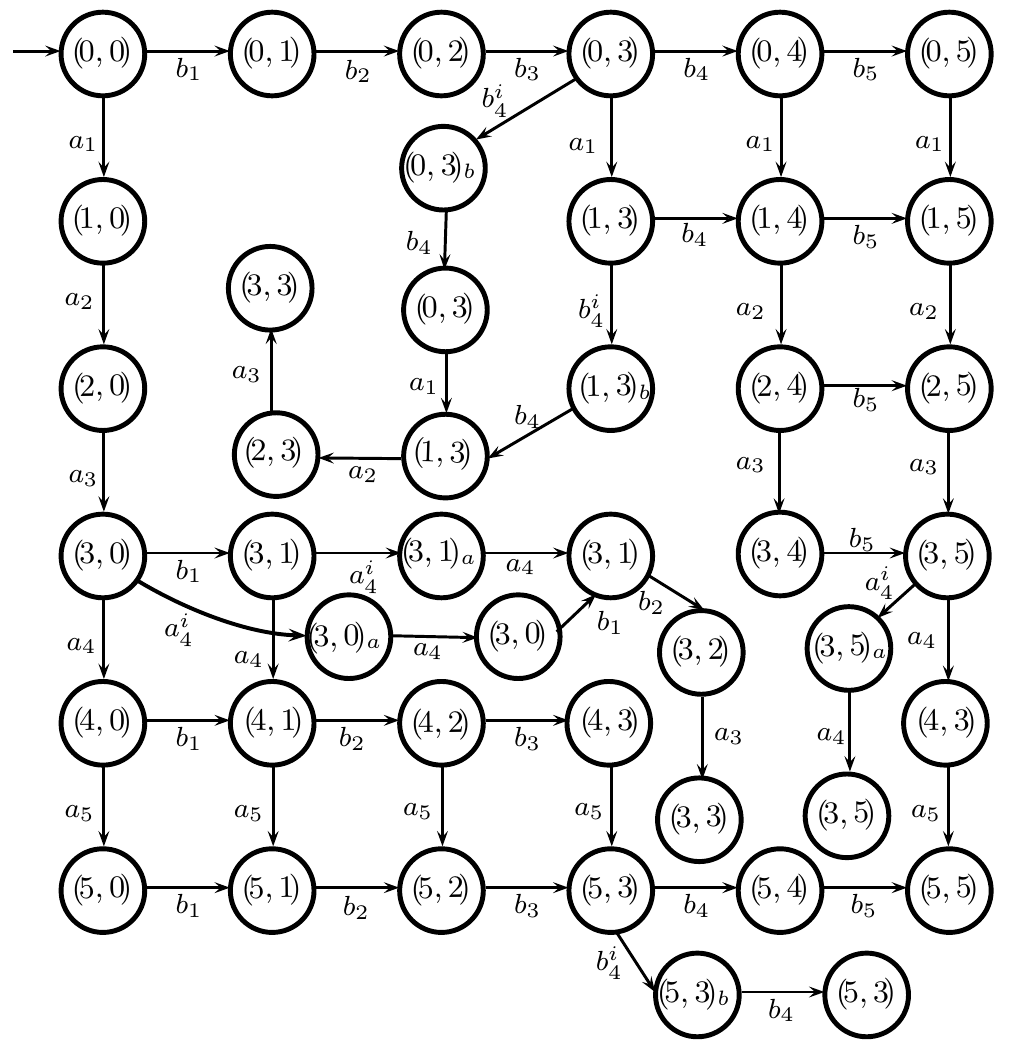}
	\caption{\label{fig:guideway_GmSI} The closed-loop system $G_M$ under SI-attack. The state labels only show the state of plant $G$.}
\end{figure}

module cannot detect the SI-attack; for example, trace $b_1b_2b_3b_4^ib_4 a_1 a_2a_3$ has the same projection as $b_1b_2b_3b_4 a_1 a_2a_3$. 
Therefore, the system is not SI-safe controllable.


\section{Conclusion} 
\label{sec:conclusion}
We have considered the problem of intrusion detection and mitigation in supervisory control systems, where the attacker can either enable or disable vulnerable actuator events and can erase or insert vulnerable sensor events. 
We presented a general methodology for modeling and analysis of  the cases of AE-, SE- and SI-attacks.
To prevent damage from  attacks, we proposed a mechanism to detect them online; upon detection, we considered the conservative approach of switching to a safe mode of operation, where all controllable events are disabled. We defined the properties of AE-, SE- and SI-safe controllability that the system should satisfy in order to successfully prevent damage from AE-, SE- and SI-attacks, respectively. 
These three properties were cast as three cases of the general form of safe-controllability, GF-safe controllability.
We developed two algorithms to test  whether a system is AE-safe controllable or not, using diagnoser or verifier automata. 
We generalized the diagnoser-based algorithm to the case of GF-safe controllablilty. 

Many problems are of interest for future investigations, such as, more detailed analyses of combinations of attacks, different attack, or different information structure for the attacker.
Extensions to other response modes from the supervisor are certainly worthy of investigation. The case where damage can be prevented after detection of an attack when the supervisor does not disable all controllable events can be approached as a supervisory control problem for a modified specification.
Such types of supervisory control problems have been considered in other published works, e.g., on multi-modal control problems \citep{Faraut:2009} and the so-called reconfiguration supervisor \citep{Nooruldeen:2015}.




\bibliographystyle{elsarticle-harv}
\bibliography{automatica_attackFINAL-v1}

\appendix
\section{AE-attack exemple}
\label{sec:exempleAEAttack}

We will now analyze AE-safe controllability. To this end, we consider the same traffic control system as in Section~\ref{sec:exemple}.
We assume that the set of vulnerable actuator events is $\Ecv=\{a_2,b_2\}\subseteq\Ec$, and thus, the set of attacked actuator events is $\Ecva=\{a_2^a,b_2^a\}$.

Following Algorithm \ref{alg:ae}, we construct, in step~1, automaton $G_a$ that models the plant under AE-attacks which is obtained by adding transitions labeled with events in $\Ecva=\{a_2^a,b_2^a\}$ in parallel with the transitions labeled with the corresponding events in $\Ecv=\{a_2,b_2\}$. In step 2, we build the realization of the supervisor under AE-attacks by adding a self-loop at every state for each event in $\Ecva$ and $\Euca=\{a_3,b_3,a_5,b_5\}$.
Finally, in step 3, we construct the closed-loop system $G_M$ under attacks, as depicted in Figure~\ref{fig:guideway_HAG}.
We can see that, after attack $a_2^a$, the closed-loop system may reach unsafe states such as state $(3,3)$. 
Next, with $G_M$ being built, we use Algorithm \ref{alg:sc} to test if the system is AE-safe controllable.
\begin{figure}[htp]
	\centering
	\includegraphics[width=0.45\textwidth]{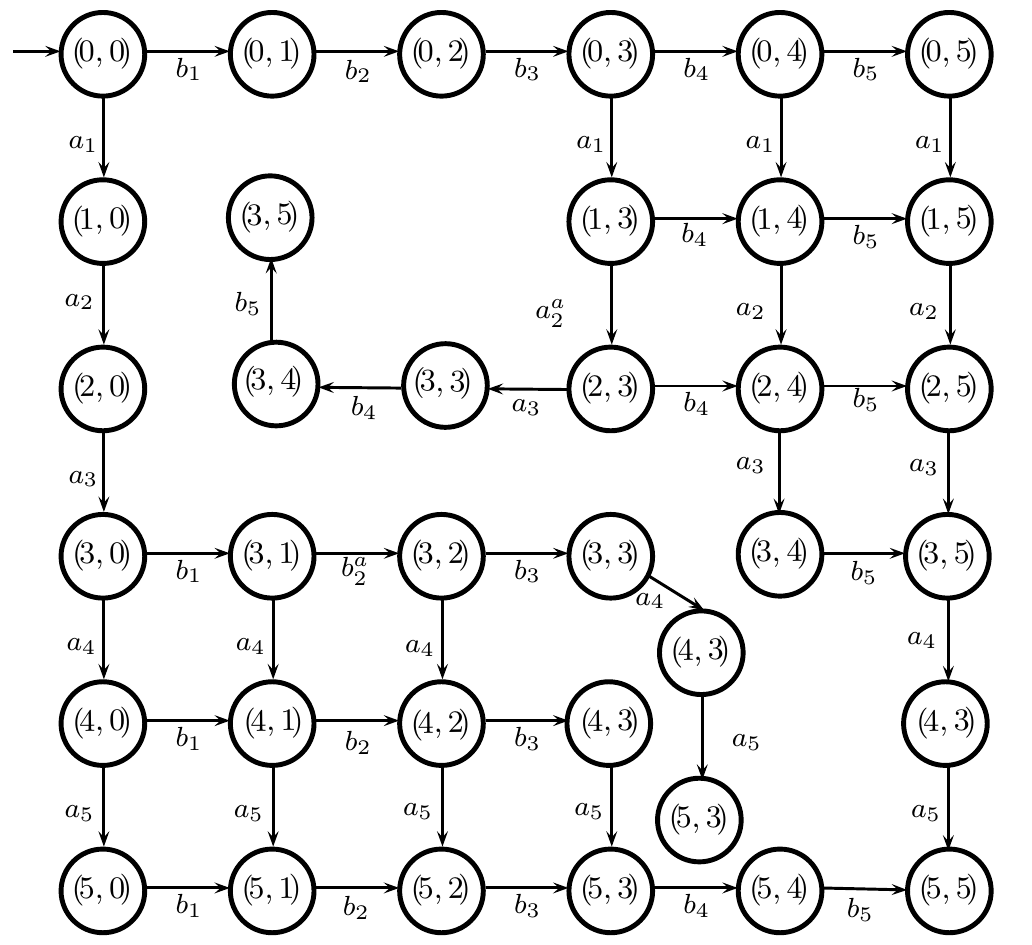}
	\caption{\label{fig:guideway_HAG} The closed-loop system $G_M= H_{a}\|G_a$ under attack. The state labels only show the state of $G_a$.}
\end{figure}
We show in Figure~\ref{fig:guideway_Gd} part of the diagnoser  with respect to $\Ecva$ built in accordance with Algorithm \ref{alg:sc}. 
\begin{figure}[hbt]
	\centering
	\includegraphics[width=0.65\linewidth]{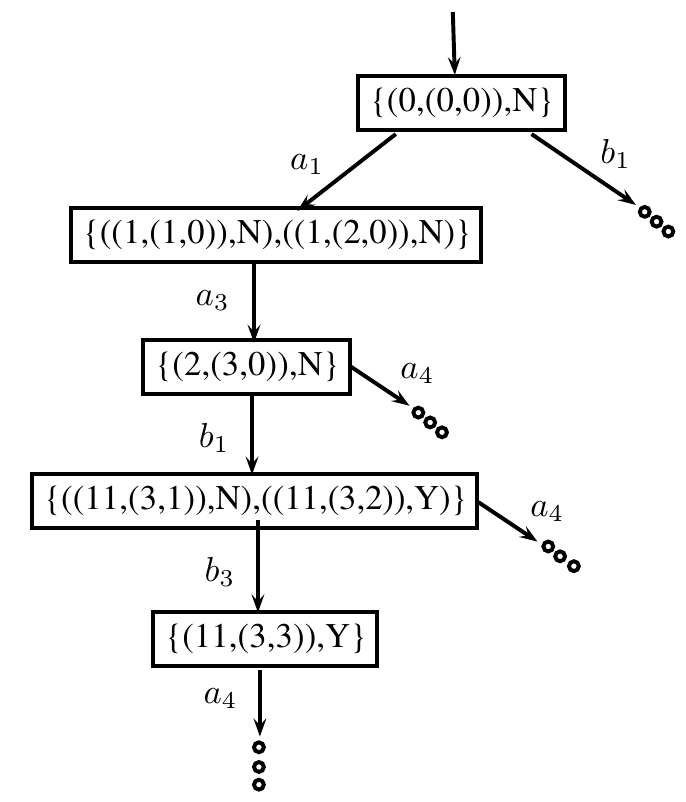}
	\caption{\label{fig:guideway_Gd} A part of diagnoser $G_{d}$.} 
\end{figure}
It can be seen in Figure~\ref{fig:guideway_Gd} that a string $s$ with projection $\Po^a(s)=a_1a_3b_1$ reaches uncertain diagnoser state $\{((11,(3,1)),N), ((11,(3,2)),Y)\}$, which surely uncontrollably reaches unsafe state $\{((11,(3,3)),Y)\}$ via event $b_3 \in \Euc$.  
Hence, the supervisor cannot prevent the plant from reaching an unsafe state after the attack is detected.
The system is not AE-safe controllable. 
\section{Proof of Proposition \ref{thm:diag} }
\begin{proof} 
	Let us denote the conclusion clauses in Definition~\ref{def:safcon} by $SC1 := (\nexists \omega \in L_M)[\Po^a(st_1)=\Po^a(\omega) \wedge \Ecva \notin \omega]$ and $SC2 := (\Ec \in t_2)$. We first prove the ``if" statement. 
	\begin{enumerate}
		\item Assume there exists $q_{YN} \in Q_{YN}$ such that $x_{i_j} \in \Xfm$ and $\ell_{i_j}=Y$ for some $ j \in \{1,\ldots ,n\}$. Then, there exists $s_d \in {\cal L}(G_d)$ such that $f_d(q_{0,d},s_d)=q_{YN}$. Thus, there exist $s_Y,s_N \in L_M$ such that $\Po^a(s_Y)=\Po^a(s_N)=s_d$, $f_M(x_{0,M},s_Y)=x_{i_j}$, $\Ecva \in s_Y$ and $\Ecva \notin s_N$. Let $s_Y=st$, where $s\in \Psi(\Ecva)$ and $t \in L_M/ s$. 
		Finally, set $\omega=s_N \in L_M$; we will have
		$\Po^a(st)=\Po^a(\omega)$ and $\Ecva \notin \omega$, which violates clause $SC1$ 
		for all decompositions of $t = t_1 t_2$.
		\item Let $q_{Y} \in \mathcal{FC}$ such that $x_{i_j} \in \Xfm$ for some $ j \in \{1,\ldots ,n\}$. Then, there exists $s_d \in {\cal L}(G_d)$ such that $f_d(q_{0,d},s_d)=q_Y$ and $s_d=s_d'\ev_o$, where $\ev_o\in \Eoa$. Thus, there exists $s_Y \in L_M$ such that $f_M(x_{0,M},s_Y)=x_{i_j}\in \Xfm$, $\Po^a(s_Y)=s_d$ and $\Ecva\in s_Y$. 
Let $s_Y=s_1s_2s_3$, where $s_1 \in \Psi(\Ecva)$, $s_2s_3 \in L_M/s_1$ and 
		$s_3 = \ev_0$.
		Taking $s = s_1 $ and $t = s_2 s_3$, we must select $t_1 = s_2 s_3$ to satisfy $SC1$, which violates $SC2$ since it forces $t_2 = \epsilon$.
		\item Let $s_Y=st_1t_2$, where $st_1=s_1s_2s_3$ as defined in case 2) to enforce $SC1$. Due to the definition of $X^{uc}$ in Algorithm \ref{alg:sc}, $t_2 \in (\Euc\cup\Ecva)^*\cap L_M/st_1$. Then, $\Eca \not\in t_2$, which violates clause $SC2$ when we enforce $SC1$.
		That is, there is no decomposition of $t = t_1 t_2$ that satisfies both $SC1$ and $SC2$.
	\end{enumerate} 
	We now prove the ``only if" part by contrapositive. 	
	Suppose that statements $1$, $2$ and $3$ are all false. This implies that {\it (i)} there does not exist an uncertain state $q_{YN}=\{(x_{i_1},\ell_{i_1}), \ldots, (x_{i_n},\ell_{i_n})\}\in Q_d $ such that $x_{i_j} \in \Xfm$ and {\it (ii)} there does not exist a first entered certain state $q_{Y}=\{(x_{i_1},Y), \ldots, (x_{i_n},Y)\}\in \mathcal{FC}$ such that $x_{i_j} \in \Xfm$; and {\it (iii)} there does not exist a state $x_M \in X^{uc}$ such that $x_M \in \Xfm$. 
	Taken together, statements {\it (i)}, {\it (ii)}, and {\it (iii)} imply that either the antecedent of $SC1 \wedge SC2$ in Definition~\ref{def:safcon} is false, or, if it is true, then both $SC1$ and $SC2$ are true for some decomposition of $t$ as $t= t_1 t_2$. Hence, $L_M$ is AE-safe controllable.
\end{proof}
\section{Proof of Proposition \ref{thm:ver}}
\begin{proof}We consider here the same clauses $SC1$ and $SC2$ of Definition~\ref{def:safcon} as in the proof of Proposition~\ref{thm:diag}. The set of events of the automata in Algorithm \ref{alg:scv} are shown in Table \ref{tab:setEvents} to help to following the proof.\newline	
	($\Leftarrow$)
	\begin{enumerate}
		\item This part is similar to case 1 of Proposition~\ref{thm:diag}. If there exists $x_V=\{(x_N,N),(x,Y)\} \in X_V$ such that $x \in \Xfm$, then, there exists $s_V \in L(G_V)$ such that $f_V(x_{0,V},s_V)=x_V$. Thus, there exist $s_Y \in L(G_F)$ and $s_N\in L(G_N)$ such that $P(s_V)=s_Y$, $P_R(s_V)=s_N$, $f_F(x_{0,F},s_Y)=(x,Y)$, $\Ef \in s_Y$ and $\Ef \not\in s_N$, where $P: (\Ea \cup \Er)^*\rightarrow \Ea^*$ and $P_R: (\Ea \cup \Er)^*\rightarrow \Er^*$. Since $G_F$ and $\tilde{G}_N$ (defined in Algorithm 1 in \citep{Moreira:2011}) 
		are subautomata of $G_M$ and $G_N$ is obtained from $\tilde{G}_N$ by renaming the unobservable events, then $s_Y \in L_M$, $s_N=R(\tilde{s}_N)$ and $\tilde{s}_N \in L_M$. Finally, setting $s_Y=st \in L_M$ and $\omega=\tilde{s}_N$, then by the construction of $G_V$, $\Po^a(st)=\Po^a(\omega)$, which violates clause $SC1$
				for all decompositions of $t = t_1 t_2$.
						
		\item This case follows the same ideas as cases 2 and 3 of Proposition~\ref{thm:diag}. If there exists $x_T=\{x^{cd}_V,(x,\ell)\} \in X_T$ such that $x_V^{cd}=A$ and $x \in \Xfm$, then, there exists $s_T\in L(G_T)$  such that $f_T(x_{0,T},s_T)=x_T$. Thus, there exists $s^{cd}_V \in L(G^{cd}_V)$ and $s_Y \in L(G_F)$ such that
		$s_T \in \{s^{cd}_V\}\cap P^{-1}(s_Y)$, where $P: (\Ea \cup \Er)^*\rightarrow \Ea^*$. By construction of automaton $G^{cd}_V$, $s^{cd}_V=s_V\ev_os_{uc}$ where $s_V \in L(G_V)$, $\ev_o \in \Eoa$ and $s_{uc} \in (\Euc\cup \Ecva)^*$. 
		Since $s_V \in L(G_V)$, thus, there exists $s'_Y=s_1s_2 \in L_M$ such that $s_1 \in \Psi(\Ef)$, $s_1s_2 \in P[L(G_V)]$ and $P(s_V)=s'_Y$. 
		Let $s_V\ev_o \in L(G_V)\Eoa\cap P^{-1}[L(G_F)]$ and define $s=s_1$ and set $t_1=s_2\ev_o$
		in order to satisfy $SC1$. 
		Since $s_T \in \{s_V\ev_os_{uc}\}\cap P^{-1}(s_Y)$, {\it i.e.}, $s_T$ tracks the trace $s_Y$ that has after $s_V\ev_o$ only uncontrollable events, thus, $t_2=s_{uc}$, which violates clause $SC2$. 
	\end{enumerate}
	($\Rightarrow$) {\it Proof by contrapositive}. Suppose the statements 1 and 2 are both false. This implies that: {\it (i)} there does not exist $x_V=\{(x_N,N),(x,Y)\} \in X_V$ such that $x \in \Xfm$; and {\it(ii)} there does not exist $\{x^{cd}_V,(x,Y)\} \in X_T$ such that $x_V^{cd}=A$ and $x \in \Xfm$. Statement {\it(i)} and {\it(ii)} imply that either the antecedent in  Definition~\ref{def:safcon} is false, or if it is true, then both $SC1$ and $SC2$ are true, where $t_1$ is chosen to match fault detection (when $A$ is entered). Hence, $L_M$ is safe controllable.
	
	\begin{table}
		\caption{Set of events of automata in Algorithm \ref{alg:scv}}
		\label{tab:setEvents}
		\begin{center}
			\begin{tabular}{cc}
				\hline
				Automaton & Set of events\\
				\hline
				$G_m$ & $\Ea$\\
				$G_F$ & $\Ea$\\
				$G_N$ & $E_N$\\
				$G_V$ & $\Ea\cup\Er$\\
				$G_V^{cd}$ & $\Ea\cup\Er$\\
				$G_T$ & $\Ea\cup\Er$\\
				\hline
			\end{tabular}
		\end{center}
	\end{table}
\end{proof}
\section{Proof of Proposition \ref{thm:mattacks}}
\proof 
Let $H_a$ be the automaton obtained in Algorithm~\ref{alg:ae} representing the all-out attacker.
The associated closed-loop language is $L_{AA}={\cal L}(H_a)\cap {\cal L}(G_a)$. 
Assume that $H'_a$ is the automaton that models the attacker that does not attack at all times.
Note that $H'_a$ can be represented as a subautomaton of $H_a$ subject to state splitting.
Therefore, we have ${\cal L}(H'_a)\subseteq {\cal L}(H_a)$, which implies that $L_{SA}={\cal L}(H'_a)\cap {\cal L}(G_a)\subseteq {\cal L}(H_a)\cap {\cal L}(G_a)=L_{AA}$.

To prove by contradiction, assume that $L_{AA}$ is AE-safe controllable but $L_{SA}$ is not AE-safe controllable.
Then, there exists a string $st \in L_{SA}$ that reaches an unsafe state, where the last event of $s$ is in $\Ecva$ and $\forall t_1,t_2$ such that $t=t_1t_2$ we do not have both $E_c\in t_2$ and ($\nexists\omega\in L_M$)($\Po(st_1)=\Po(\omega)\wedge\Ecva \not\in \omega$). 
Because $L_{SA} \subseteq L_{AA}$, the same $st$ is also in $L_{AA}$ contradicting the AE-safe controllability of $L_{AA}$. 
Therefore, $L_{SA}$ must be AE-safe controllable.
\endproof

\end{document}